\documentclass[prb,twocolumn,superscriptaddress,showpacs,floatfix,floats]{revtex4-1}
\usepackage{textcomp,graphicx,natbib}

\newcommand{\texp}[1]{\ensuremath{\times 10^{#1}}}
\newcommand{\dgC}{\ensuremath{^\circ\mbox{C}}}

\begin{document}
\title{Electromagnon in ferrimagnetic $\varepsilon \mbox{-Fe}_2\mbox{O}_3$ nanograin ceramics}
\affiliation{Institute of Physics, Academy of Sciences of the Czech Republic, Na Slovance 2, 182 21 Prague~8, Czech Republic}
\affiliation{Institut de Ci\`encia de Materials de Barcelona---Consejo Superior de Investigaciones Cient\'\i ficas, Campus UAB, 08193, Bellaterra, Catalunya, Spain}
\affiliation{Institut Laue-Langevin, BP 156, 38042 Grenoble Cedex 9, France}
\affiliation{Faculty of Mathematics and Physics, Department of Condensed Matter Physics, Charles University, Ke Karlovu 5, 121 16 Prague 2, Czech Republic}
\affiliation{Grenoble High Magnetic Field Lab, CNRS - 25, avenue des Martyrs, Grenoble Cedex 9, France}
\author{Christelle Kadlec}
\affiliation{Institute of Physics, Academy of Sciences of the Czech Republic, Na Slovance 2, 182 21 Prague~8, Czech Republic}
\author{Filip Kadlec}
\thanks{Corresponding author}
\email[e-mail: ]{kadlecf@fzu.cz}
\affiliation{Institute of Physics, Academy of Sciences of the Czech Republic, Na Slovance 2, 182 21 Prague~8, Czech Republic}
\author{Veronica Goian}
\affiliation{Institute of Physics, Academy of Sciences of the Czech Republic, Na Slovance 2, 182 21 Prague~8, Czech Republic}
\author{Mart\'{i} Gich}
\affiliation{Institut de Ci\`encia de Materials de Barcelona---Consejo Superior de Investigaciones Cient\'\i ficas, Campus UAB, 08193, Bellaterra, Catalunya, Spain}
\author{Martin Kempa}
\affiliation{Institute of Physics, Academy of Sciences of the Czech Republic, Na Slovance 2, 182 21 Prague~8, Czech Republic}
\author{St\'ephane Rols}
\affiliation{Institut Laue-Langevin, BP 156, 38042 Grenoble Cedex 9, France}
\author{Maxim Savinov}
\affiliation{Institute of Physics, Academy of Sciences of the Czech Republic, Na Slovance 2, 182 21 Prague~8, Czech Republic}
\author{Jan Prokle\v{s}ka}
\affiliation{Faculty of Mathematics and Physics, Department of Condensed Matter Physics, Charles University, Ke Karlovu 5, 121 16 Prague 2, Czech Republic}
\author{Milan Orlita}
\affiliation{Grenoble High Magnetic Field Lab, CNRS - 25, avenue des Martyrs, Grenoble Cedex 9, France}
\author{Stanislav Kamba}
\affiliation{Institute of Physics, Academy of Sciences of the Czech Republic, Na Slovance 2, 182 21 Prague~8, Czech Republic}

\begin{abstract}
Electromagnons are known from multiferroics as spin waves excited by the 
electric component of electromagnetic radiation. We report the discovery of an 
excitation in the far-infrared spectra of $\varepsilon 
\mbox{-Fe}_2\mbox{O}_3$ which we attribute to an 
electromagnon appearing below 110\,K, where the ferrimagnetic 
structure becomes incommensurately modulated.  Inelastic neutron scattering 
shows that the electromagnon energy corresponds to that of a magnon from the 
Brillouin zone boundary.  Dielectric measurements did not reveal any 
sign of
ferroelectricity in $\varepsilon \mbox{-Fe}_2\mbox{O}_3$ down to 10\,K, 
despite its acentric  crystal structure. This shows that the 
activation of an electromagnon requires, in addition to the polar 
ferrimagnetic structure, a modulation of the magnetic structure. We 
demonstrate that a combination of inelastic neutron scattering with infrared and 
/ or terahertz spectroscopies allows detecting electromagnons in ceramics, where no 
crystal-orientation analysis of THz and infrared spectra is possible.
\end{abstract}
\pacs{76.50.+g, 77.22.-d, 63.20.kd, 75.85.+t}
\maketitle

\section{Introduction}
In the last years, there has been an increasing interest in so-called 
multiferroic
materials, displaying simultaneously spontaneous ferroelectric (FE) polarization 
and ferro- or
 antiferromagnetic (AFM) ordering. Multiferroics exhibit a rich variety 
of fundamental
physical phenomena, and it is generally believed that they have a
potential for novel applications in non-volatile memories \cite{Scott07,Roy12},
magnonics~\cite{Kruglyak10} and magnetic sensors \cite{Nan08}. These 
	applications would rely on the coupling of order parameters on various time 
	scales, from quasi-static to ultrafast. However, the understanding of the 
	microscopic mechanism of the magnetodielectric coupling is still
a fundamental problem of solid state physics. The static and dynamic 
magnetoelectric (ME) couplings can have different origins. Owing to the 
static ME coupling, the macroscopic FE polarization emerges in the cycloidal 
or transverse conical modulated magnetic structures; this polarization can 
change with magnetic field. In contrast, the dynamic ME coupling generates an 
oscillatory polarization and leads to a dielectric dispersion in the terahertz 
(THz) region. Indeed, THz studies of
multiferroics revealed a new kind of electric-field-active spin excitations
contributing to the dielectric permittivity $\varepsilon = 
\varepsilon^{\prime}-\mbox{i}\varepsilon^{\prime\prime}$, called
electromagnons (EMs)~\cite{Pimenov06}. Their characteristic feature is a 
coupling with polar phonons, which
manifests itself in the spectra by a transfer of dielectric strength from 
phonons to
EMs on cooling \cite{ValdesAguilar07}. In contrast to ferromagnetic and AFM
resonances, which are magnons from the Brillouin zone (BZ) center contributing 
to the
magnetic permeability $\mu= \mu^{\prime}-\mbox{i}\mu^{\prime\prime}$, the EMs
can be activated also outside of the BZ center
\cite{ValdesAguilar09,Takahashi12,Stenberg12,Mochizuki10}. The 
understanding of this fact is not trivial, because the photons which excite EMs 
have wavevectors much smaller than the EMs. Thus, to date, there are several 
different theories attempting to explain the observed properties of EMs in 
various materials \cite{ValdesAguilar09,Stenberg12,Mochizuki10,Khomskii09}.

The EMs were discovered first in TbMnO$_3$ and GdMnO$_3$ \cite{Pimenov06} 
which belong to multiferroics denoted~\cite{Khomskii09} as type II, where the FE 
order is induced by a special magnetic ordering. Since then, EMs were confirmed 
in numerous type-II 
multiferroics~\cite{ValdesAguilar09,ValdesAguilar07,Sushkov07,Sushkov08,Pimenov08,Kida09,Seki10,Kezsmarki11,Shuvaev11}.  
Other reports of EMs in type-I multiferroics (e.g.  
BiFeO$_{3}$~\cite{Cazayous08,Talbayev11,Komandin10} or hex-YMnO$_3$ 
\cite{Pailhes09}) appear inconclusive, since no transfer of the dielectric 
strength from polar phonons to EMs was observed~\cite{Cazayous08,Pailhes09}.  
Also, recent infrared IR and THz studies did not confirm the EM in hex-YMnO$_3$ 
\cite{Kadlec11}.

Here we report experiments which reveal an excitation identified as an EM in 
the  ferrimagnetic $\varepsilon$ phase of $\rm Fe_2O_3$. Thanks to 
its chemical simplicity, this phase appears also as a suitable model system for 
theoretical studies of electromagnonic excitations. While $\varepsilon$-$\rm 
Fe_2O_3$ is quite rare and less known than the $\alpha$ (hematite) or
$\gamma$ (maghemite) phases of $\varepsilon \mbox{-Fe}_2\mbox{O}_3$~\cite{Machala11}, its
properties make it attractive for applications, such as electromagnetic-wave absorbers
and memories~\cite{Namai08,Namai12,Ohkoshi07}. Owing to limited phase stability, it can
be synthesized only in the form of nanoparticles tens of nanometers in size
\cite{Namai12,Tucek10}, epitaxial thin films~\cite{Gich10} or nanowires a few micrometers long \cite{Ding07}. 
Below 480--495\,K, it is ferrimagnetic~\cite{Jin04,Sakurai05}; at room-temperature, it 
has a collinear spin structure~\cite{Tucek11} and exhibits a coercive
field of $H_{\rm c}\approx 2\,\mbox{T}$~\cite{Jin04}---the highest known value among
metal oxides. The crystal lattice has a temperature-independent 
non-centrosymmetric orthorhombic structure with the $Pna2_1$ space group 
\cite{Tronc98} (magnetic space group $Pn^{\prime}a2^{\prime}_1$). It consists of 
three crystallographically non-equivalent
$\rm FeO_6$ octahedra, forming chains along the $a$ direction, and one type of $\rm
FeO_4$ tetrahedra~\cite{Tucek10,Gich06}. Compared to isostructural GaFeO$_{3}$, the
low-temperature phase diagram of $\varepsilon$-$\rm Fe_2O_3$ is complex---below 
150\,K, a series of magnetic phase transitions occurs. Below $T_{\rm m}=110\,\mbox{K}$, an incommensurate 
magnetic ordering appears where the magnetic structure
modulation has a periodicity of about 10
unit cells \cite{Gich06}. Near
$T_{\rm m}$, a drop in $\varepsilon^{\prime}$ was observed, and magnetocapacitive
measurements revealed a quadratic coupling \cite{Gich06a}. 
Room-temperature microwave measurements provided evidence of a strong ferromagnetic
resonance (FMR) near 0.74\,\mbox{meV} (frequency of 180\,GHz) which can be tuned by
doping with Al, Ga or Rh~\cite{Namai08,Ohkoshi07,Namai12}. In order to gain insight into
the dynamic ME properties of $\varepsilon$-$\rm Fe_2O_3$, we obtained
THz, IR and inelastic neutron scattering (INS) spectra of $\varepsilon$-$\rm Fe_2O_3$ nano-grain ceramics
upon cooling down to 10\,K, providing information about polar and magnetic excitations.

\section{Samples and experimental methods}
The nanoparticles of $\varepsilon$-$\rm Fe_2O_3$ were synthesized by sol-gel 
chemistry. $\rm SiO_2$-$\rm Fe_2O_3$ composite gels containing 30 wt.\,\% of $\rm 
Fe_2O_3$ were prepared from iron nitrate nonahydrate (Sigma-Aldrich $>98\%$) and 
tetra\-ethoxy\-silane (TEOS, Sigma-Aldrich 98\%) in hydroethanolic medium at 
TEOS:H$_2$O:EtOH = 1:6:6 molar ratio. Iron nitrate was first dissolved and then 
TEOS added dropwise to the mixture under stirring. The sol was poured into 5\,cm 
diameter petri dishes that were closed with its cover and gelation took place 
for between 4 and 5 weeks. The gels were dried overnight in a stove at 70\,\dgC, 
crushed and thermally treated in air atmosphere for 3 hours at 1100\,\dgC\ 
(heating rate 80\,\dgC/h). The resulting material was a composite of 
$\varepsilon$-$\rm Fe_2O_3$ nanoparticles of about 25\,nm in diameter dispersed 
in an amorphous $\rm SiO_2$ matrix as checked by X-ray diffraction (XRD) which 
did not reveal any trace of other $\rm Fe_2O_3$ polymorphs. The silica was 
removed by stirring the composite powder for 12\,h in a 12M aqueous NaOH 
solution at 80\,\dgC\ under reflux. XRD patterns recorded after the silica 
removal revealed that the microstructure and the phase stability of 
$\varepsilon$-$\rm Fe_2O_3$ nanoparticles were not affected by the etching 
process. The nanoparticles were further processed by spark plasma sintering 
(SPS) in order to prepare a pellet suitable for dielectric, terahertz (THz)  and 
IR measurements by pressing the 
$\varepsilon$-$\rm Fe_2O_3$ powder in a graphite mould for 4 minutes at 
350\,\dgC\ under 100\,MPa. The XRD analysis of the sintered pellet showed that 
the SPS process did not induce any grain growth or phase transformation. Finally, the SPS pellets were polished to thin disks with a thickness of 1.2\,mm. Some 
IR and THz measurements were performed on $\varepsilon$-$\rm Fe_2O_3$ pellets 
with a diameter of about 6\,mm, which were prepared from powder at room 
temperature using a standard tabletop manual hydraulic press (Perkin Elmer). The 
spectra were qualitatively the same, only the value of the high-frequency IR 
reflectance was affected by the roughness of the sample surface, which could not 
be polished.

IR reflectance measurements with the resolution of 0.25\,meV were
performed using the Fourier transform infrared spectrometer Bruker
IFS-113v in near-normal reflectance geometry with an incidence angle of
$11^{\circ}$. An Oxford Instruments Optistat optical cryostat with
polyethylene windows was used for sample cooling down to 10\,K, and a 
liquid-He-cooled Si
bolometer operating at 1.6\,K was applied as a detector. We also measured far-IR 
reflectivity with applied magnetic field up to 13\,T. To this aim,
	another Bruker IFS-113v spectrometer and a custom-made      superconducting 
	magnetic cryostat allowing the measurements at 2 and 4\,K were used. 
	Time-domain THz
spectroscopy was based on measurements of sample transmittance using custom-made
spectrometers based on Ti:sapphire femtosecond lasers; one with an Optistat
cryostat with mylar windows for measurements without magnetic field but with a
higher frequency resolution, enabling to discern the FMR profile, and one
with an Oxford Instruments Spectromag cryostat, enabling measurements with 
magnetic
field of up to 7\,T. Here, the Voigt configuration was used with the external 
static magnetic field
$B_{\rm ext}$ perpendicular to the magnetic component of the THz radiation 
$B_{\rm THz}$.  Similar
effects were observed also for $B_{\rm ext}\parallel B_{\rm THz}$.

INS experiments were performed between 10 and 190\,K using about 3\,g of 
loose $\varepsilon \mbox{-Fe}_2\mbox{O}_3$ nanopowder in the IN4
time-of-flight diffractometer at the Institut Laue-Langevin in Grenoble, France.

\section{Results and discussion}
\subsection{Broad-band study of the electromagnetic response.}
\begin{figure}
	\raggedright
	\hspace*{2.5pt}\includegraphics[width=0.867\columnwidth]{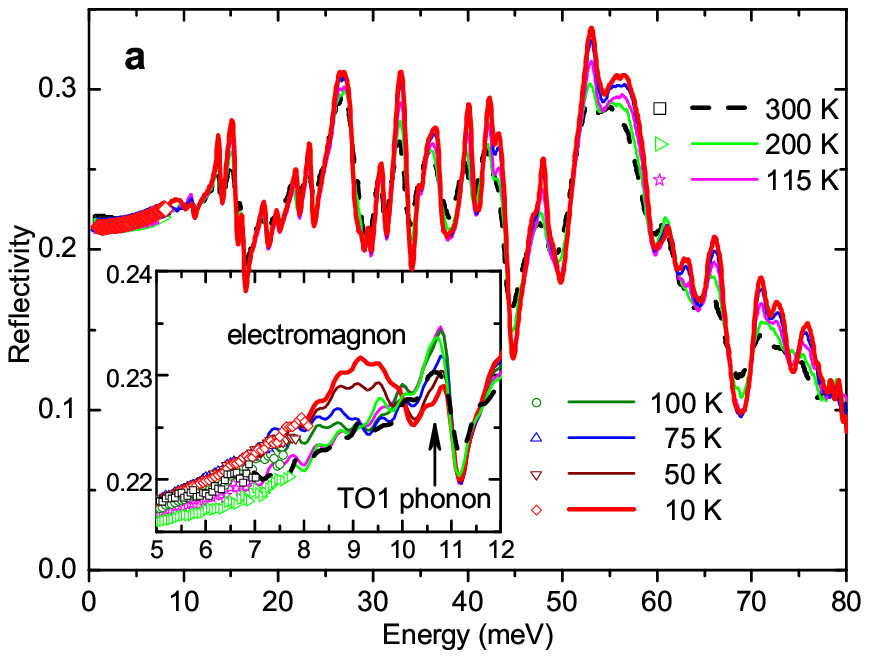}
	\includegraphics{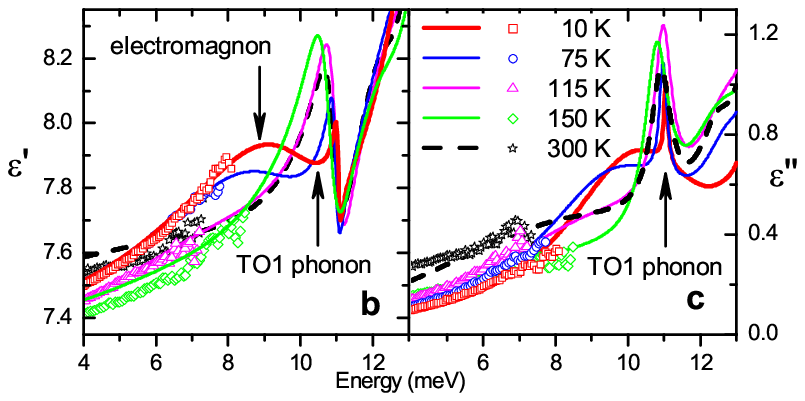}
	\caption{(a) Lines: IR reflectivity spectra showing polar phonons. Symbols 
		below 8\,\mbox{meV}: data calculated from THz spectra. The
		inset shows in detail the low-energy part where, below 100\,K and
		10\,\mbox{meV}, a new reflection band appears due to the EM. (b), (c): 
		Fits of the complex permittivity in the far IR
		region, obtained from the IR reflectivity spectra using a sum of 
		harmonic oscillators
		(lines), compared to data obtained from  THz spectroscopy
	(symbols).}
	\label{fig:IRtemp}
\end{figure}
\begin{figure}
	  \centering
	  \includegraphics[width=0.75\columnwidth]{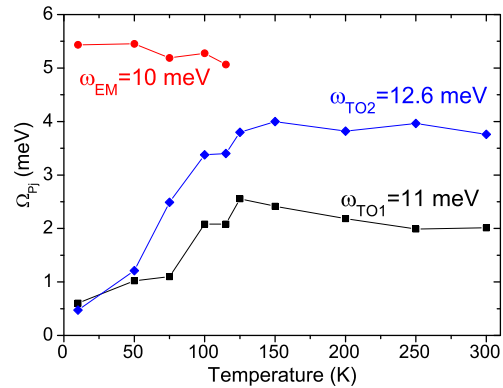}
	  \caption{Temperature dependence of the plasma frequencies 
		  (defined as $\Omega_{\rm pj}=\sqrt{\Delta\varepsilon_j}\omega_j$) of the 
	  10-meV-mode attributed to EM and of the TO1 and TO2 phonons. The 
  dielectric strengths $\Delta\varepsilon_j$ were evaluated by fitting using a 
  model with harmonic oscillators.}
	  \label{fig:transfer}
  \end{figure}
Fig.\ \ref{fig:IRtemp}a shows the far and mid-IR reflectivity spectra 
displaying polar optical phonons
of $\varepsilon$-$\rm Fe_2O_3$ between 10 and 300\,K. Figs.\ \ref{fig:IRtemp}b, c 
	show the far-IR $\varepsilon(E)$ spectra calculated from 
	the fits of IR reflectivity together with the experimental THz data.
	To this purpose, we used a model involving 35 harmonic oscillators; 
		this number is lower than the number of IR active modes provided by the 
		factor group analysis (see Appendix A); apparently, a part of the modes 
	are too weak to be observed. Upon cooling, all phonons above 12\,meV
	exhibit the usual behavior---their intensity increases due to reduced
	phonon damping at low temperatures. The TO1 phonon near 11\,meV exhibits an
	anomalous behavior: on cooling, its intensity increases only down to
	115\,K. Below this temperature, it markedly weakens,
	while a supplementary broad reflectivity peak develops below
	$E\sim 10\,\mbox{meV}$ and becomes more intense upon cooling (see the inset of Fig.\ \ref{fig:IRtemp}a).
	This transfer of strengths involves also the TO2 phonon (see Fig.\ 
		\ref{fig:transfer}), evidencing a coupling  
	among these three polar modes. Despite the lattice distortions 
	which occur between 150\,K and 75\,K, the crystal symmetry of $\varepsilon 
	\mbox{-Fe}_2\mbox{O}_3$ does not change with temperature 
	\cite{Gich06,Tseng09}. This is further confirmed by our IR reflectivity
	spectra, displaying a temperature-independent number of polar phonons; 
	should a structural phase transition occur, it would imply a change 
	of the 
factor group analysis and different phonon selection rules. Given the high 
number of atoms in the unit cell, multiple new reflection bands throughout the 
IR range would be observed. Therefore, one can exclude the new mode to originate 
in a structural modification.

Another option to be considered is the polar phonon splitting due to 
exchange coupling below AFM phase transitions which was reported in various 
transition-metal monoxides and chromium spinels \cite{Kant12}; the mode 
splitting increased on cooling below the N\'eel temperature. However, this 
explanation cannot be valid as we observe an 
opposite temperature dependence---the new mode appears below $T_{\rm m}$ at low 
energies and hardens towards the TO1 phonon energy on cooling, i.e.\ their 
energy difference decreases. 

Finally, one cannot a priori exclude the hypothesis of activation of the 
	TO1 phonon branch from the area of the BZ near its edge. This would require 
	a
folding of the structural BZ which could be caused by a transfer of the magnetic 
BZ folding (linked to incommensurability) via magnetostriction.  Nevertheless, 
in the X-ray diffraction studies, no appropriate satellite reflections were 
observed. Even supposing these satellite reflections to be very weak, one would 
expect the off-center phonons to activate also at higher energies, which we did 
not observe. This hypothesis therefore seems unlikely. Based on further 
experimental evidence, especially in view of an analogous temperature 
behavior observed by INS, we argue below that the reflection band 
activated below $T_{\rm m}$ is most probably an EM.

 \begin{figure}
     \centering
\includegraphics[width=0.51\columnwidth]{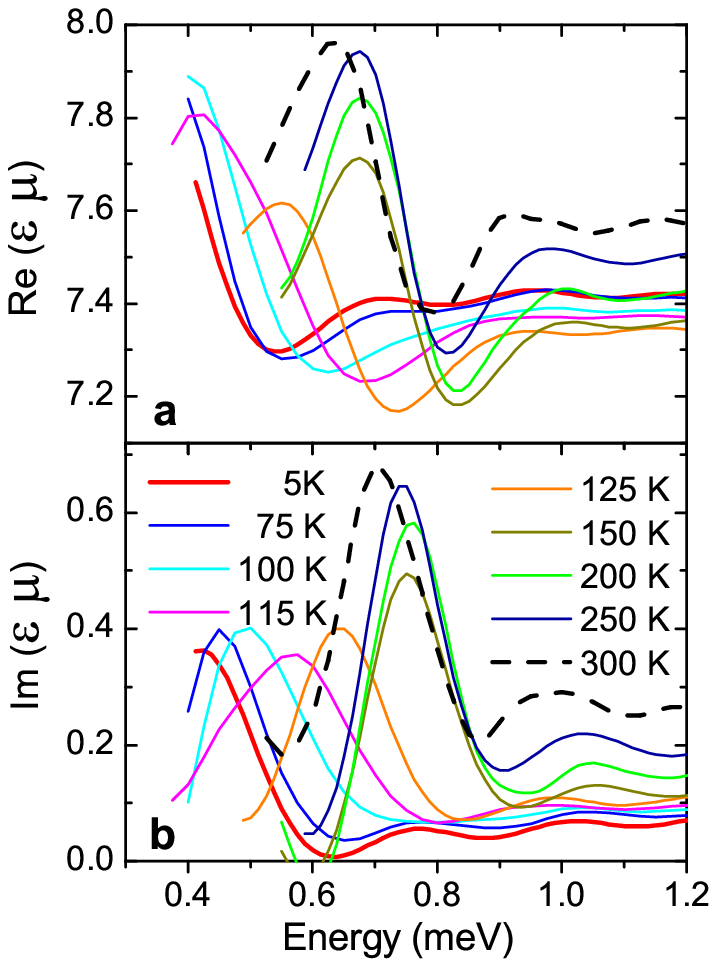}%
\begin{minipage}[b]{0.48\columnwidth}%
\includegraphics[width=\textwidth]{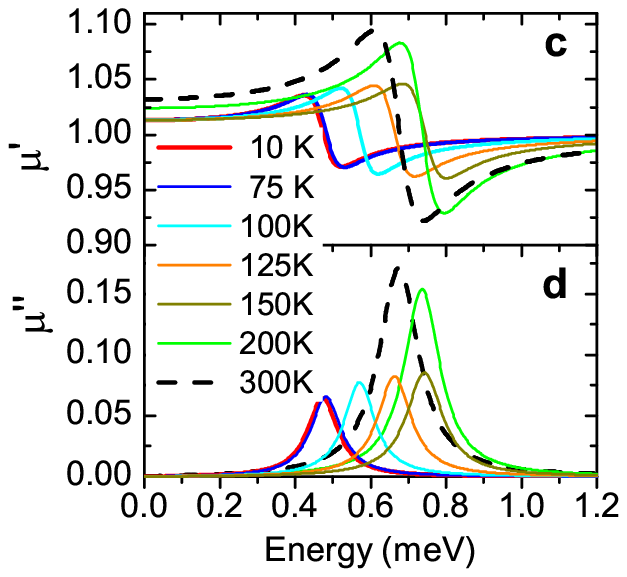}\\
\includegraphics[width=\textwidth]{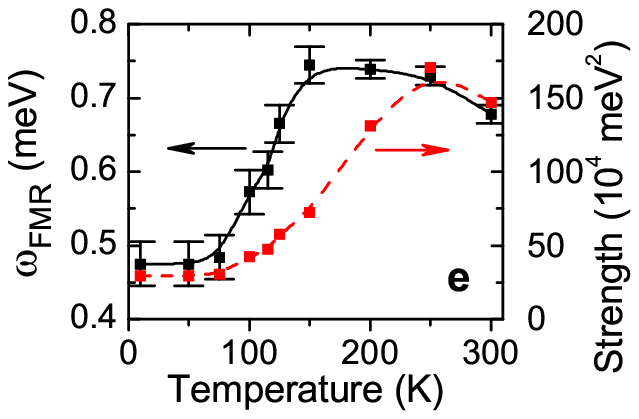}
\end{minipage}
    \caption{Temperature dependence of the spectra of the (a) real and (b)
    imaginary parts of the $\varepsilon\mu$ product, obtained by
    THz  spectroscopy. Spectra of $\mu^{\prime}$ (c),
    $\mu^{\prime\prime}$ (d), corresponding to the FMR mode, obtained by fitting the THz
    spectra. (e) Temperature dependence of the FMR
    energy and strength $\Delta\mu \omega_{\rm FMR}^{2}$ derived from parts c, d.}
    \label{fig:magnonT}
\end{figure}

\begin{figure*}[ht!]
	\centering
        \includegraphics[width=0.48\textwidth]{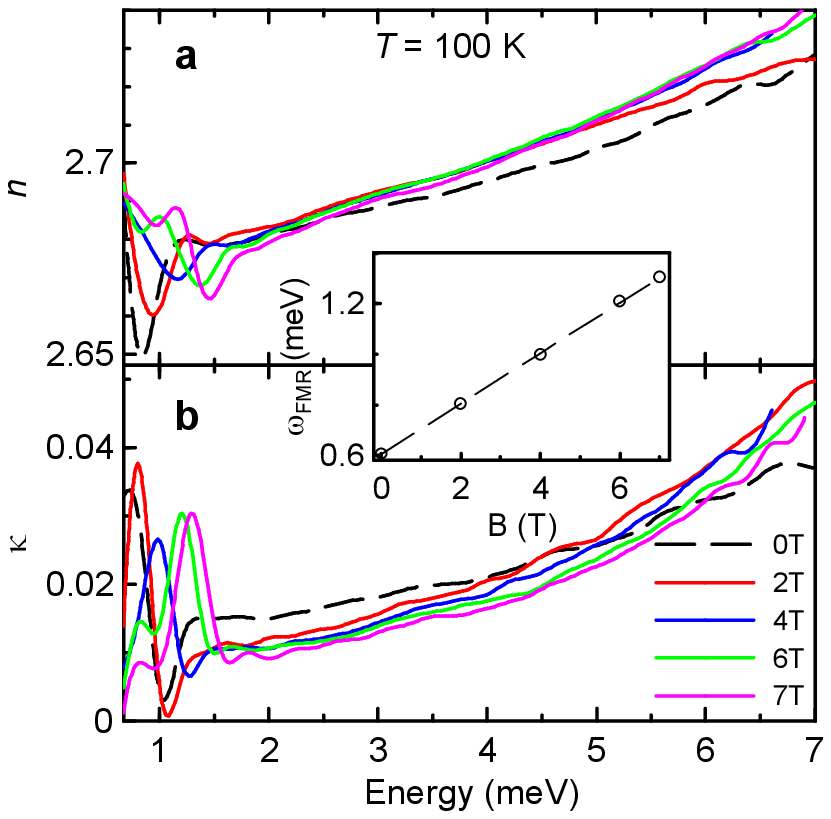}
        \includegraphics[width=0.48\textwidth]{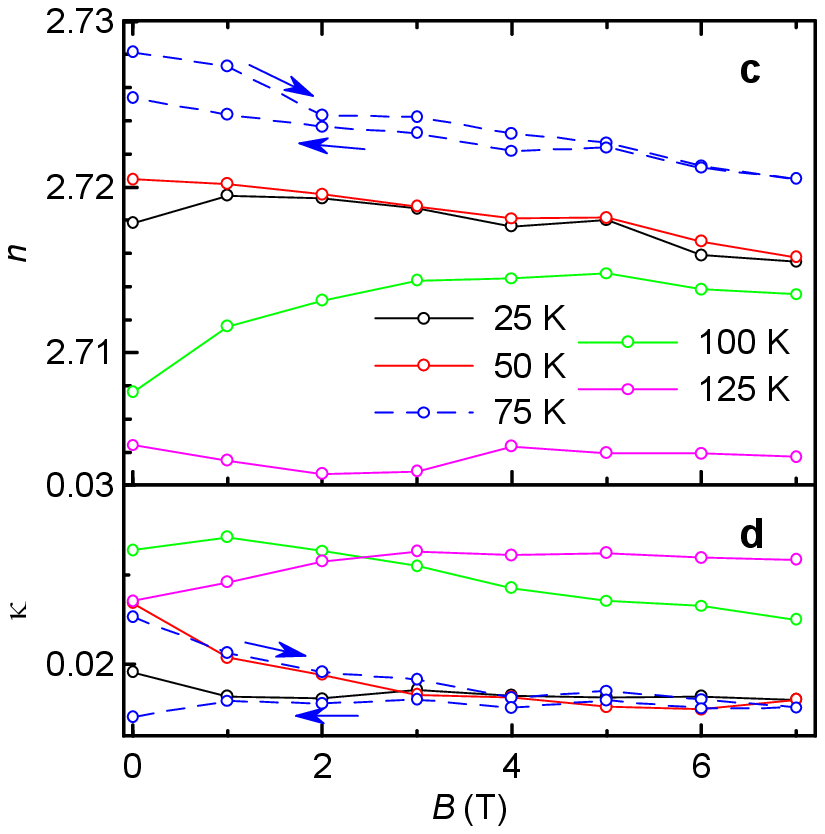}
    \caption{(a), (b): Spectra of complex refractive index $N\equiv
        n-\mbox{i}\kappa$ of $\varepsilon \mbox{-Fe}_2\mbox{O}_3$ measured by
        THz spectroscopy at
        $T=100\,\mbox{K}$ as a function of applied magnetic field. Inset: 
		$B$-dependence of the FMR frequency, determined as the peak in 
		$\kappa(E)$ spectra. (c), (d):
        Changes of the value of $n$, $\kappa$, determined within
        $\pm0.001$, for $E=5\,\mbox{meV}$  as a function of temperature and
    increasing magnetic field (except at 75\,K).
    \label{fig:NkTHz100K}}
\end{figure*}
The temperature dependent THz spectra  (see Fig.  \ref{fig:magnonT}) reveal the 
sharp FMR which was previously reported at room 
temperature~\cite{Namai08,Namai12}. To quantify its temperature behavior, we 
used the harmonic oscillator model for all phonons and one term accounting for 
the FMR in $\mu(E)$, while assuming a smooth dependence of $\varepsilon(E)$ in 
this interval. The resulting spectra, matching well the measured data, are shown 
in  Fig.\ \ref{fig:magnonT}c,d. From the fit parameters, we derived the 
temperature dependence of the magnon strength and FMR  energy (see  Fig.\ 
\ref{fig:magnonT}e).  We observe a sharp drop  in the resonance energy between 
150\,K and 75\,K, very similar to that of the coercive field $H_{\rm 
c}(T)$~\cite{Gich05}. This can be explained by the fact that the FMR energy is 
proportional to the magnetocrystalline anisotropy field $H_a$.  As the sample 
consists of randomly oriented particles with a uniaxial magnetic anisotropy, 
$H_a$ is proportional to the $H_{\rm c}$ value \cite{Ohkoshi07}.

\begin{figure}[h]
	\centering
		\includegraphics[width=0.42\textwidth]{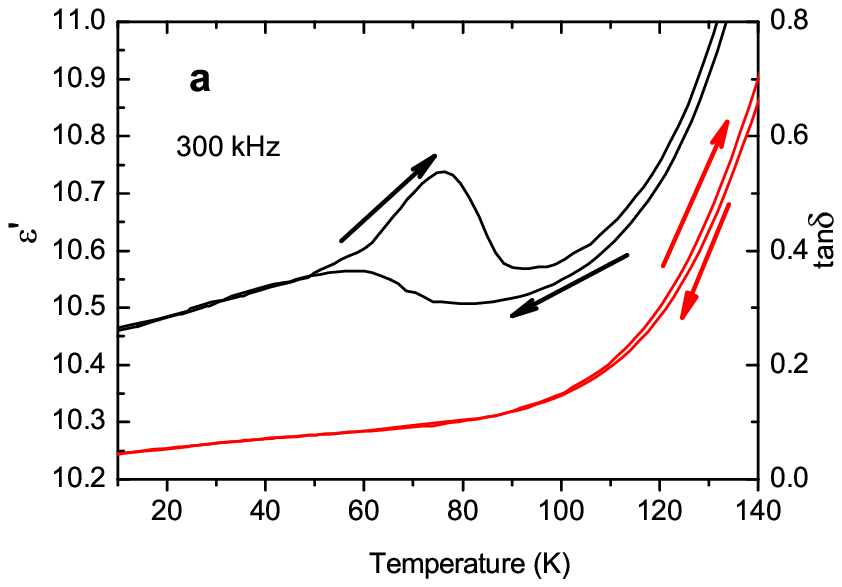}\\
		\includegraphics[width=0.38\textwidth]{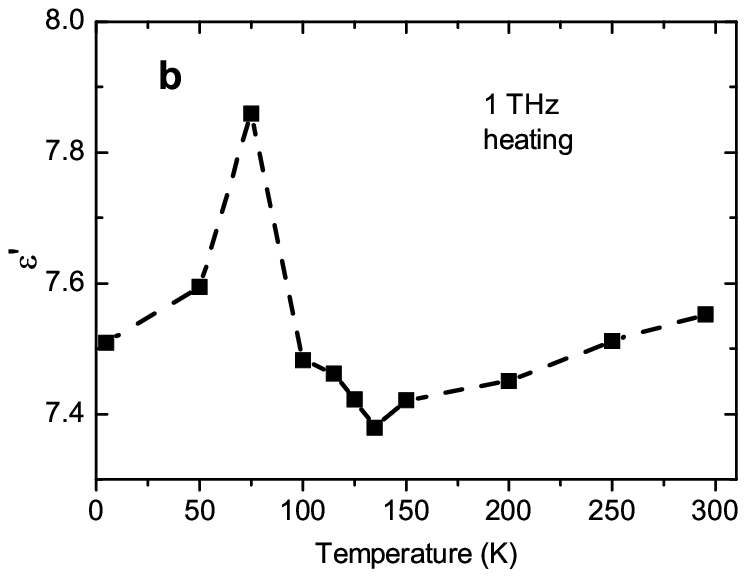}\\
		\includegraphics[width=0.4\textwidth]{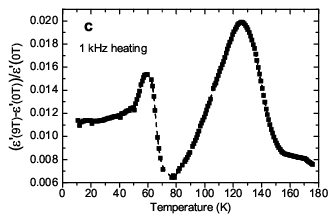}
	\caption{(a) Temperature hysteresis of the dielectric permittivity 
		(black lines, left 
		axis) and
		losses (red lines, right axis) observed at 300\,kHz. (b) Temperature dependence of the permittivity at
	 1\,THz measured on heating. The dashed line is a guide to the eyes. The
	 values at 300\,kHz are systematically higher than at 1\,THz due to a small
	 dielectric relaxation between
	 these two frequencies; one can see a similar permittivity peak near 75\,K 
	 in
 both experiments. (c) Temperature dependence 
		of relative changes of the 1\,kHz-permittivity due to magnetic field with $B= 
		9\,\mbox{T}$ (taken on heating).}
	\label{fig:SupImped}
\end{figure}

Furthermore, we measured THz time-domain spectra with external magnetic field 
ranging from 0 to 7\,T. Because of the high absorption of the EM, lying near 10 
meV, the sample was opaque above 7\,meV. Therefore, we could measure only the 
low-frequency wing of the EM. When the magnetic field is applied, two types of 
changes in the THz spectra can be observed: an increase of the FMR frequency 
corresponding to the peak of the $\kappa(E)$ spectra, and a change of the slope 
of both real and imaginary parts of the index of refraction, indicating shifts 
of the EM frequency with magnetic field. An example of the former behavior at 
$T=100\,\mbox{K}$ is shown in Fig.~\ref{fig:NkTHz100K}a, b; the FMR frequency, 
upon applying a static magnetic field of $B=7\,\rm T$, increases from 0.6 to 
1.3\,\mbox{meV} (see inset of Fig.~\ref{fig:NkTHz100K}a, b). The latter 
phenomenon is illustrated by Fig.~\ref{fig:NkTHz100K}c, d which traces the 
values of the complex refractive index at $E=5\,\mbox{meV}$ as a function of 
temperature and applied magnetic field. While changes only close to the 
sensitivity level were detected at temperatures of 10 and 300\,K (not shown in 
Fig.~\ref{fig:NkTHz100K}), there is a clear $B$-dependence of the spectra at 
intermediate temperatures. The highest sensitivity was observed at 100\,K, close 
to the magnetic phase transition.  Also, at $T=75\,\mbox{K}$, a marked 
hysteresis in $B$ occurs, similarly to the temperature hysteresis observed by 
radio-frequency impedance spectroscopy techniques near this temperature (see 
Figure~\ref{fig:SupImped}); this observation will be discussed below. At 
$T\ll T_{\rm m}$, where the magnetic structure is probably stable, the 
changes of $N$ with magnetic field are smaller. This explains also why we did 
not detect any significant changes of the far-IR spectra with magnetic field at 
$T=2$\,K.

\begin{figure}[t]
	\includegraphics[width=\columnwidth]{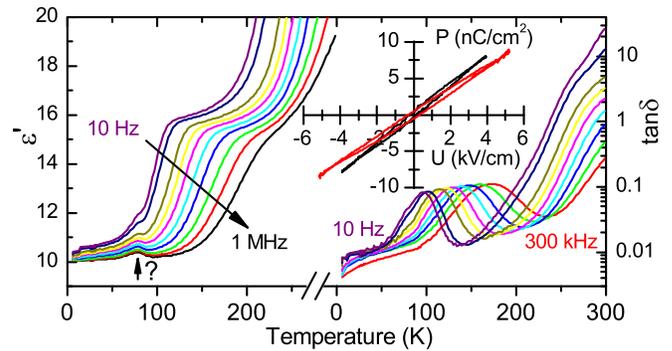}
		\caption{Temperature dependence of the real permittivity
        $\varepsilon^\prime$ (left) and dielectric losses $\tan\delta$ (right),
		measured upon heating by impedance spectroscopy.  Inset:
		dependence of the polarization on the applied 50\,Hz ac bias at 120\,K 
	(black) and 15\,K (red).}\label{fig:kHz}
\end{figure}

In the frequency range from $f=$10\,Hz to 1\,MHz, the complex permittivity 
$\varepsilon$ was
measured by impedance spectroscopy as a function of temperature (see Fig.\ \ref{fig:kHz}).
No sign of a FE phase transition was detected. Above 200\,K, both $\varepsilon^{\prime}(T)$ and
$\varepsilon^{\prime\prime}(T)$ increase due to the leakage
conductivity and the related  Maxwell-Wagner polarization. Between 100 and
200\,K, we observed a step-like decrease of $\varepsilon^{\prime}(T)$ towards lower
temperatures and maxima in losses
$\tan\delta(T,f)=\varepsilon^{\prime\prime}(T,f)/\varepsilon^{\prime}(T,f)$, which is typical
of a dielectric relaxation.  The temperature dependence of the relaxation time
$\tau(T)$ obtained from the peaks of $\tan\delta(T,f)$  
follows an Arrhenius
behavior, $\tau(T)=\tau_{0}\mbox{e}^{{E_0}/{k_{\rm B}T}}$ with $k_{\rm B}$ denoting the
Boltzmann constant, $\tau_{0}=(1.5\pm 0.2)\texp{-12}\,\mbox{s}$ and $E_0=(0.195\pm
0.002)\,\mbox{eV}$. The origin of this relaxation is not clear, however, similar effects
are known from several perovskite rare-earth manganites, including the multiferroics
TbMnO$_3$ and DyMnO$_3$~\cite{Schrettle09}. We attribute the relaxation 
to thermally activated vibrations of the FE domain walls or magnetic domain walls which can be 
polar \cite{Pyatakov11}. The huge room-temperature coercive field $H_{\rm c}$ is
the consequence of a single-domain magnetic structure of the 
nanograins~\cite{Namai12}. Below 200\,K, $H_{\rm c}$ strongly decreases 
due to a transition to a polydomain structure \cite{Gich05} which explains why 
the dielectric relaxation exists only in this temperature range.

The inset of Fig.\ \ref{fig:kHz} shows the measured dependences of the 
polarization on applied electric field.  No open FE hysteresis loops nor signs 
of saturation were observed under the applied fields. Since the $Pna2_1$ crystal structure of 
$\varepsilon \mbox{-Fe}_2\mbox{O}_3$ corresponds to a pyroelectric space group, 
we cannot exclude that an applied electric field with an intensity higher than 
the one we used (beyond 5\,kV/cm, our sample became leaky) would switch the 
polarization and that $\varepsilon \mbox{-Fe}_2\mbox{O}_3$ is in fact
FE. Actually, one of us recently investigated strained epitaxial $\varepsilon 
\mbox{-Fe}_2\mbox{O}_3$ thin films and, under an applied electric field one 
order of magnitude stronger, observed a room-temperature FE 
switching.\cite{Gich-prep} Since the crystal symmetry of $\varepsilon 
\mbox{-Fe}_2\mbox{O}_3$ does not change with temperature \cite{Nizn-priv}, 
one can not exclude that the $\varepsilon \mbox{-Fe}_2\mbox{O}_3$ nanograins are 
also FE already above the ferrimagnetic phase transition occurring near 490\,K; 
in any case, it is at least pyroelectric. Consequently, $\varepsilon 
\mbox{-Fe}_2\mbox{O}_3$ would belong to type-I multiferroics.

Near 75\,K, a small peak in $\varepsilon^{\prime}(T)$ was observed in our 
impedance spectroscopy measurements (as marked by 
the arrow in Fig.\ \ref{fig:kHz}). This peak is rather weak on cooling, but it 
becomes more distinct on heating, and it exhibits a temperature hysteresis of 
$\approx$15\,K (see also Fig.~\ref{fig:SupImped}). This is reminiscent of a 
dielectric anomaly typical for pseudoproper or improper FE phase transitions, 
such as those in perovskite rare-earth manganites. However, this hypothesis is 
not confirmed by the polarization measurements shown in Fig.\ 
\ref{fig:kHz}, and the X-ray and neutron diffraction investigations did not 
reveal any structural changes near 75\,K either \cite{Gich06,Tseng09}. In 
type-II multiferroics, a narrow dielectric peak is seen at $T_{\rm c}$ only at 
frequencies below 1\,MHz and its intensity strongly decreases with rising 
frequency \cite{Schrettle09}. By contrast, in our impedance spectra, the 
peak is present at all frequencies up to the THz region (see 
Fig.~\ref{fig:SupImped}b), although it is partly covered by the stronger 
dielectric relaxation at low frequencies.  Therefore, this anomaly must 
originate from phonons or an EM. As the observed dielectric anomaly occurs at a 
temperature close to the lowest-temperature magnetic phase transition 
\cite{Gich06}, we propose that it arises from the transfer of the dielectric 
strength from the TO1 and TO2 phonons to the EM (see Fig.\ 
	\ref{fig:transfer}).  We note that in single-crystal 
multiferroics, often a step-like increase of the permittivity occurs below the 
temperature where the electromagnon activates.~\cite{Sushkov08} Our observations 
on nanograin samples are somewhat different---while a step-like increase of 
$\varepsilon^{\prime}$ below $\approx130$\,K, superimposed with the narrow-range anomaly near 75\,K, 
was detected in the THz range (see Fig.~\ref{fig:SupImped}b), only the anomaly 
near 75\,K manifests itself in the kHz range (see Fig.~\ref{fig:SupImped}a). We 
suppose that the step in the low-frequency permittivity is screened by the observed 
dielectric relaxation in the microwave range.

We also investigated the dependence of the permittivity at 1\,kHz on 
	external magnetic field up to 9\,T. We found that $\varepsilon'(B)$ exhibits 
	the highest changes (almost 2\%) near 70 and 130\,K (see Fig.~\ref{fig:SupImped}c). Both 
	of these anomalies are clearly linked to the changes of magnetic 
	structure~\cite{Gich06}. We 
	suppose that the lower-temperature change
	corresponds to the EM anomaly observed also in THz experiments, while that
	observed near 130\,K is due to the relaxation linked to the magnetic and 
	simultaneously polar domain walls.

\begin{figure}
        \includegraphics[width=0.485\columnwidth]{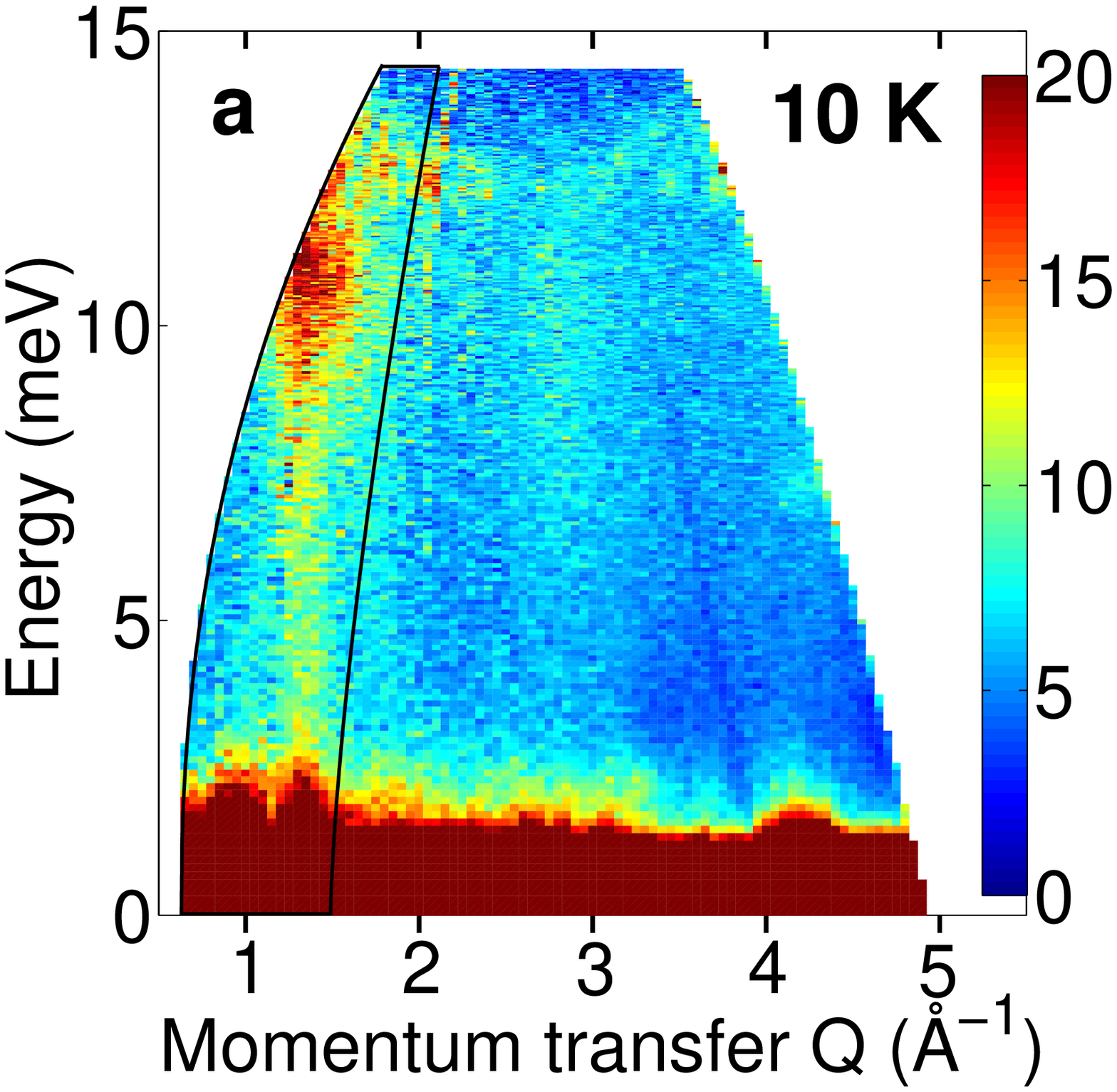}
        \includegraphics[width=0.445\columnwidth]{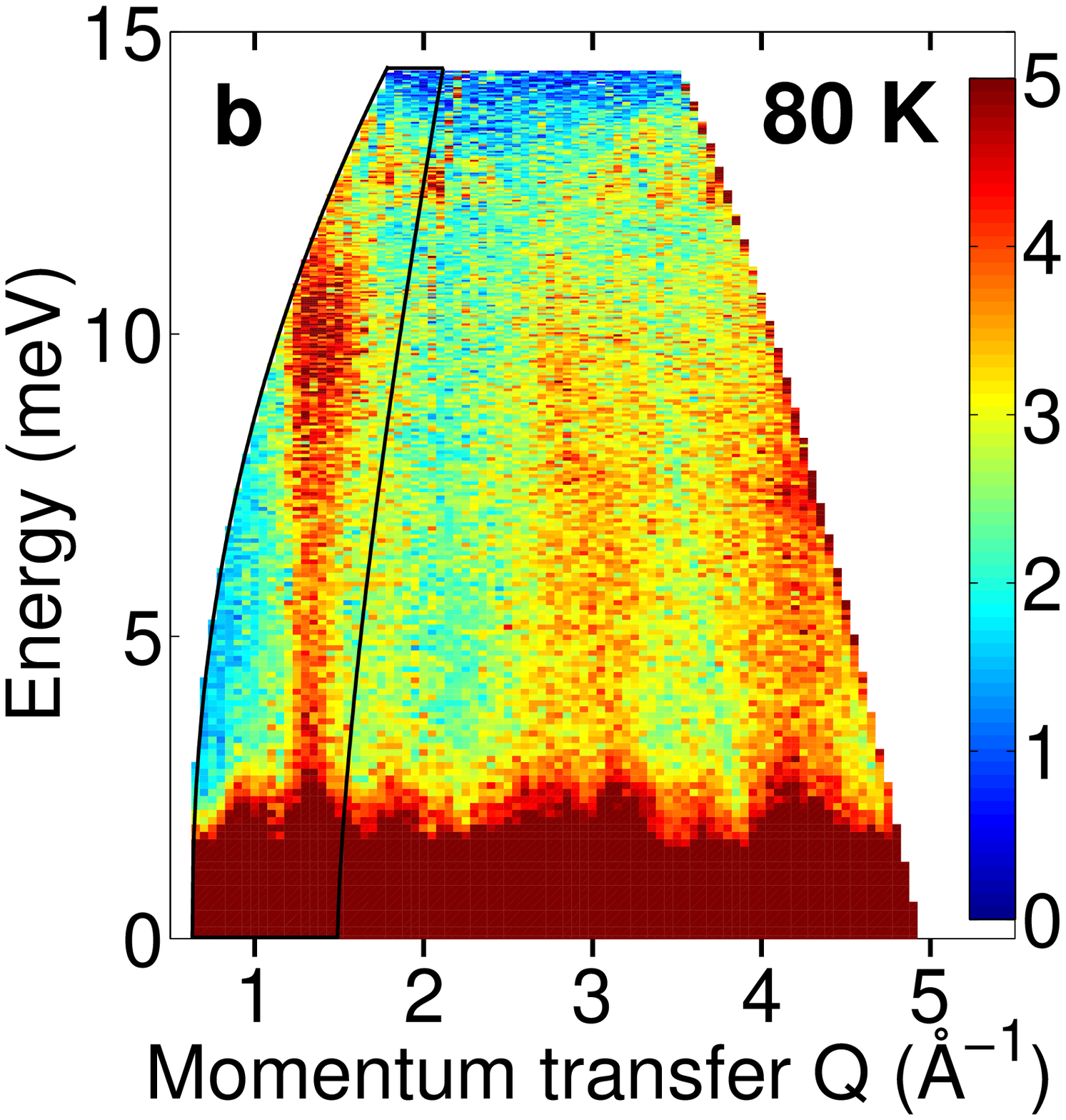}
        \includegraphics[width=0.45\columnwidth]{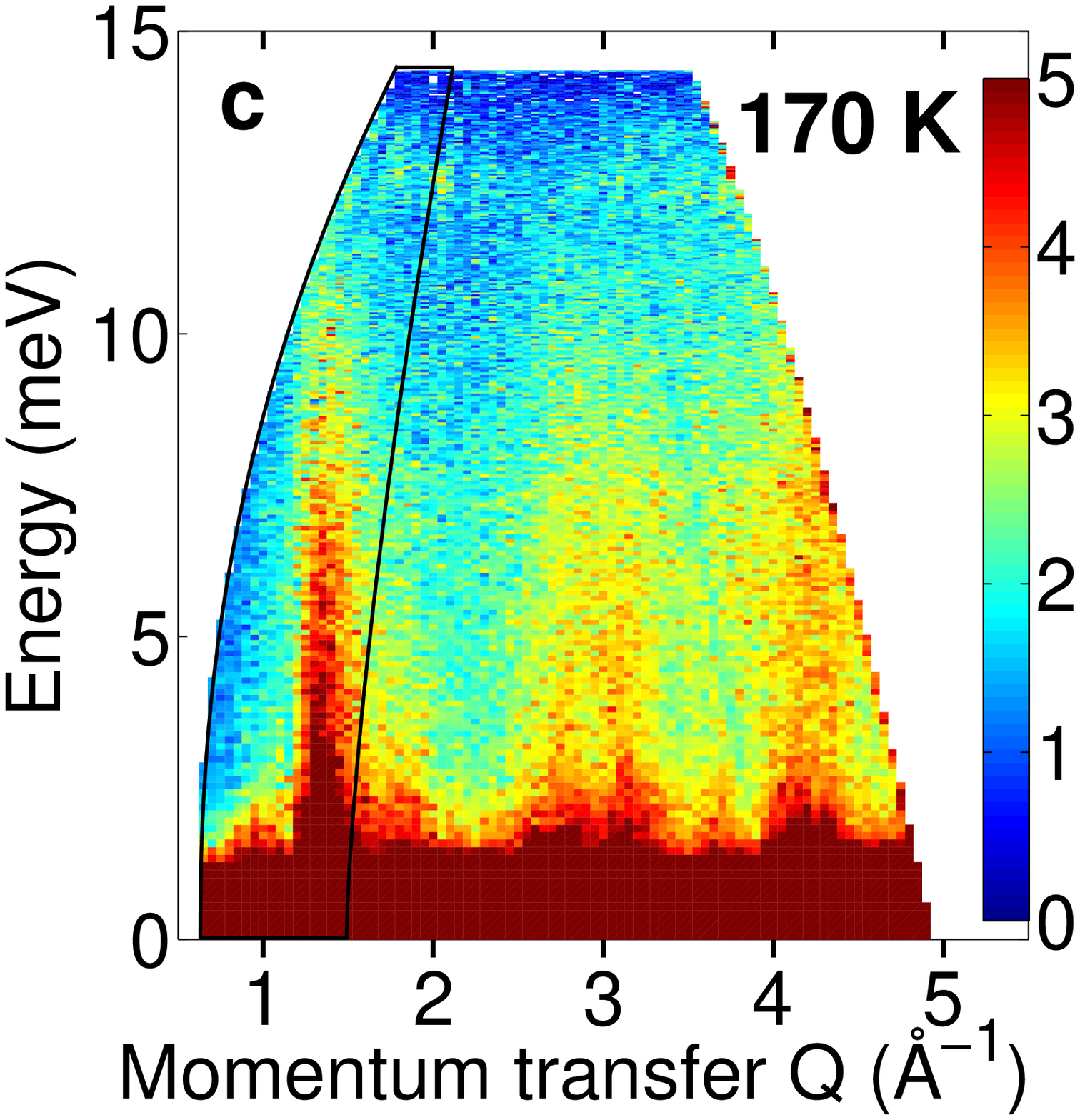}
        \includegraphics[width=0.48\columnwidth]{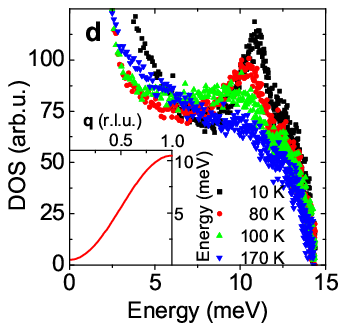}
    \caption{(a), (b), (c): Bose-Einstein-factor-normalized INS intensity as a
        function of momentum $Q$ and energy $E$ transfers for $T=10$, 80 and
        170\,K. Near $Q=1.4\,\mbox{\AA}^{-1}$, a magnon branch with a cut-off
        energy of $\approx11\,\mbox{meV}$ can be seen. (d): DOS determined by
        integrating over the regions marked by black solid lines
        in (a)--(c).  Inset of (d): scheme of the magnon dispersion branch in
    reciprocal lattice units, involving the FMR and EM near the BZ center and
boundary, respectively. \label{fig:neutron_map}}
\end{figure}

\subsection{Neutron scattering.}
In order to further explore the hypothesis of an EM, we performed time-of-flight 
INS experiments which allow measuring the phonon and magnon density of states 
(DOS) in the meV energy range. As the nanopowder does not allow us to determine 
directly the phonon and magnon dispersion branches in the BZ, the data represent 
an orientation-averaged scattering function $S(Q,E)$ where $Q$ is the total 
momentum transfer and $E$ the energy transferred between the crystal lattice and 
the neutrons (see Fig.\ \ref{fig:neutron_map}). The data reveal a steep column 
of intense scattering, emanating from magnetic Bragg peaks at $Q = 
1.4\,\mbox{\AA}^{-1}$, and extending up to $E \sim 10\,\mbox{meV}$. The weaker 
columns at $Q > 2\,\mbox{\AA}^{-1}$ are due to scattering in higher-order BZs. 
The fact that the area of most intense scattering is located at low $Q$ shows 
unambiguously \cite{Shirane06} that the dominant contribution to the low-$Q$ 
scattering comes from spin waves.

A qualitatively similar magnon response was 
recently observed in INS spectra of polycrystalline BiFeO$_{3}$ 
\cite{Delaire12}; the spin wave character of the excitation was confirmed by INS 
on BiFeO$_{3}$ crystals, where the magnon dispersion branch was directly 
measured \cite{Jeong12}. Our scattering from the magnon waves becomes weaker on 
cooling due to the decreasing Bose-Einstein factor. Around 10\,meV, a distinct 
scattering peak persists down to low temperatures, corresponding to a maximum of 
the magnon DOS; this is obviously due to a flat end of the branch below the BZ 
boundary. Moreover, the energy at the maximal magnon DOS, as well as its 
temperature evolution, corresponds to that of the newly IR-activated mode (see 
Fig.\ \ref{fig:neutron_map}d). 

The inset of Fig.\ \ref{fig:neutron_map}d shows a schematic view of an acoustic-like magnon 
dispersion branch 
giving rise to the observed excitations, both the one below 10.5\,meV (at the BZ 
boundary) and the FMR near 0.5\,meV (in the BZ center). This dispersion 
	behavior is similar to that observed in the ferrimagnetic $\rm HoFe_2$ \cite{Rhyne78}, which exhibits a 
slightly higher Curie temperature of 597\,K. In $\varepsilon 
\mbox{-Fe}_2\mbox{O}_3$, the optic-like magnon 
branches lie probably above 12\,meV, beyond the energy range used in our INS 
experiments. We suggest that this acoustic-like magnon is activated in the IR spectra due to the loss of magnetic 
translation symmetry in the incommensurate magnetic phase below $T_{\rm m}$. 
Such an activation is analogous to that of phonons with $q\neq0$ in structurally 
modulated crystals \cite{Petzelt81}. We suppose that the large damping of the 
newly activated excitation can be explained by an activation of the magnon DOS in 
the IR spectra. Since the observed spin-wave excitation is coupled with the 
lowest-energy TO1 phonon, it must be excited by the electric component of the 
electromagnetic radiation; at the same time, it has to contribute to dielectric 
permittivity. Therefore, the excitation seen near 10\,meV must be an EM.

\section{Conclusion}
In conclusion, in $\varepsilon \mbox{-Fe}_2\mbox{O}_3$, we have discovered 
an excitation, appearing simultaneously with the modulation of the magnetic 
structure, at energies below the TO1 phonon. We attribute this excitation to
an EM whose energy corresponds to a magnon from the BZ boundary.  We did not 
observe any other excitation at lower energies, in contrast to type-II multiferroics.  
There, the Dzyaloshinskii-Moriya (D.-M.) interaction breaks the center of 
symmetry, induces ferroelectricity 
\cite{Khomskii09} and the EMs are activated thanks to magnetostriction 
(($\textbf{S}_{i}\cdot\textbf{S}_{j}$)-type interaction) 
\cite{ValdesAguilar09}. In $\varepsilon \mbox{-Fe}_2\mbox{O}_3$, the crystal 
structure is acentric at all temperatures and it permits to activate the D.-M. 
interaction in an originally collinear ferrimagnetic structure~\cite{Fennie08}; 
the D.-M.\ interaction tilts the spins and finally induces an incommensurately 
modulated magnetic structure below $T_{\rm m}=110\,\mbox{K}$, where the EM activates due to 
magnetostriction.

Up to now, EMs were reported mainly in type-II multiferroics. Previous reports 
of EMs in type-I multiferroics were lacking evidence of their coupling with 
polar phonons, e.g.\ in BiFeO$_3$~\cite{Cazayous08,Talbayev11,Komandin10} or 
hex-YMnO$_3$ \cite{Pailhes09}. Our results indicate that $\varepsilon 
\mbox{-Fe}_2\mbox{O}_3$ belongs to type-I multiferroics; it is pyroelectric and 
perhaps FE even above the ferrimagnetic phase transition \cite{Nizn-priv} at 
490\,K, but the EM is activated only below $T_{\rm m}$, 
corresponding to the onset of the incommensurately modulated magnetic structure. 
In our case, a clear transfer of dielectric strength from a low-energy phonon to 
the zone boundary magnon was observed.

Finally, we would like to stress that EMs were previously identified only 
in single crystals using a thorough polarization analysis of measured spectra.  
Here we have determined an EM from unpolarized IR and THz spectra of nanograin 
ceramics showing its coupling with a TO1 phonon.  Simultaneously, we have shown 
from INS experiments made on powder that the EM in $\varepsilon$-$\rm Fe_2O_3$ 
comes from the BZ boundary. This combination of experimental methods provides a 
guideline for an unambiguous determination of EMs in materials where 
sufficiently large single crystals for polarized IR and THz measurements are not 
available.

\begin{acknowledgments}
This work was supported by the Czech Science Foundation (project\ P204/12/1163). 
The experiment in ILL Grenoble was carried out at the IN4 spectrometer 
within the project LG11024 financed by the Ministry of Education of the Czech
Republic. M.G.\ acknowledges funding from the Spanish Ministerio de Econom\'\i a y Competitividad (projects
RyC-2009-04335, MAT 2012-35324 and CONSOLIDER-Nanoselect-CSD2007-00041) and the European
Commission (FP7-Marie Curie Actions, PCIG09-GA-2011-294168). S.K.\ thanks Petr
Br\'{a}zda for his stimulation of our $\varepsilon$-$\rm Fe_2O_3$ research and 
S.\ Artyukchin for a helpful discussion.
\end{acknowledgments}

\appendix

\section{Phonons in $\varepsilon \mbox{-Fe}_2\mbox{O}_3$}
 For the orthorhombic $Pna2_1$
crystal structure of $\varepsilon\mbox{-Fe}_2\mbox{O}_3$ with 8 formula units
per unit cell \cite{Gich06}, the factor group analysis predicts the following
phonon counts and symmetries in the BZ center:
\begin{eqnarray}
\nonumber \Gamma&=30 A_{1}(z,x^{2},y^{2},z^{2}) + 30 A_{2}(xy) +\\
&+30 B_{1}(x,xz) + 30 B_{2}(y,yz).                                                             	\label{group-analysis}
\end{eqnarray}

Here, $x$, $y$ and $z$ mark electric polarizations of the IR wave for which
the phonons are IR active, while the rest of symbols are components of the Raman
tensor. After subtraction of the three acoustic phonons, 87 IR-active phonons
are expected. We have observed 35 of them (see their parameters in Table~\ref{tab:Oscil}); the remaining ones cannot be 
identified, either because of low
intensities or because they overlap with other ones.
\begin{table}[h!]
    \centering
	\renewcommand{\arraystretch}{0.57}
\begin{tabular}{|r|r@{.}l|r@{.}l|r@{.}l||r|r@{.}l|r@{.}l|r@{.}l|}
    \hline No.&
    \multicolumn{2}{|c}{$\Delta\varepsilon$}&
    \multicolumn{2}{|c}{$\Omega_0$\,[meV]$\!$}&
    \multicolumn{2}{|c||}{$\!\Gamma$\,[meV]$\!$}&No.&
    \multicolumn{2}{|c}{$\Delta\varepsilon$}&
    \multicolumn{2}{|c}{$\Omega_0$\,[meV]$\!$}&
    \multicolumn{2}{|c|}{$\!\Gamma$\,[meV]$\!$}\\
    \hline
EM &\phantom{1}0&27\phantom{1} &\phantom{1}10&47 & \phantom{1}4&67 &
						18 &\phantom{1}0&02\phantom{1} &\phantom{1} 38&40\phantom{1} &\phantom{1}1&34  \\ \hline
1 &0&01 & 11&05 & 0&13  &19 &0&18 & 40&13 & 1&40  \\ \hline
 2 &0&01 & 12&61 & 0&87  &20 &0&15 & 42&16 & 1&37  \\ \hline
 3 &0&08 & 13&85 & 0&44  &21 &0&13 & 43&37 & 1&48  \\ \hline
 4 &0&24 & 15&25 & 0&82  &22 &0&02 & 46&76 & 0&99  \\ \hline
 5 &0&06 & 16&26 & 0&49  &23 &0&16 & 48&04 & 1&87  \\ \hline
 6 &0&02 & 17&58 & 1&82  &24 &0&02 & 49&39 & 1&20  \\ \hline
 7 &0&07 & 18&64 & 0&83  &25 &0&25 & 52&78 & 2&24  \\ \hline
 8 &0&03 & 19&95 & 0&76  &26 &0&40 & 55&42 & 4&68  \\ \hline
 9 &0&09 & 21&87 & 0&99  &27 &0&07 & 57&45 & 3&03  \\ \hline
10 &0&08 & 23&42 & 0&65  &28 &0&11 & 60&84 & 3&32  \\ \hline
11 &0&56 & 27&18 & 2&33  &29 &0&03 & 63&02 & 1&95  \\ \hline
12 &0&02 & 28&88 & 0&77  &30 &0&14 & 65&66 & 3&62  \\ \hline
13 &0&01 & 29&59 & 0&46  &31 &0&07 & 70&78 & 2&29  \\ \hline
14 &0&09 & 30&91 & 0&96  &32 &0&03 & 72&58 & 1&96  \\ \hline
15 &0&21 & 33&05 & 1&12  &33 &0&07 & 75&29 & 3&61  \\ \hline
16 &0&01 & 34&94 & 0&44  &34 &0&05 & 78&46 & 4&71  \\ \hline
17 &0&37 & 36&44 & 2&60  &35 &0&08 & 85&68 & 5&43  \\ \hline
\end{tabular}
    \caption{Set of parameters used in the oscillator model to fit the IR
    reflectance data at 10\,K. $\Delta\varepsilon$, $\Omega_0$ and $\Gamma$ mark 
	the dielectric contribution, eigenfrequency and damping of polar modes. The first row contains the parameters of the electromagnon, 
the other rows describe 35 polar phonons. From mid-IR reflectivity, 
the high-frequency electronic contribution was obtained as 
$\varepsilon_{\infty}=3.2.$}    \label{tab:Oscil}
\end{table}

\printfigures

\begin{thebibliography}{49}
\expandafter\ifx\csname natexlab\endcsname\relax\def\natexlab#1{#1}\fi
\expandafter\ifx\csname bibnamefont\endcsname\relax
  \def\bibnamefont#1{#1}\fi
\expandafter\ifx\csname bibfnamefont\endcsname\relax
  \def\bibfnamefont#1{#1}\fi
\expandafter\ifx\csname citenamefont\endcsname\relax
  \def\citenamefont#1{#1}\fi
\expandafter\ifx\csname url\endcsname\relax
  \def\url#1{\texttt{#1}}\fi
\expandafter\ifx\csname urlprefix\endcsname\relax\def\urlprefix{URL }\fi
\providecommand{\bibinfo}[2]{#2}
\providecommand{\eprint}[2][]{\url{#2}}

\bibitem[{\citenamefont{Scott}(2007)}]{Scott07}
\bibinfo{author}{\bibfnamefont{J.}~\bibnamefont{Scott}}, \bibinfo{journal}{Nat.
  Mater.} \textbf{\bibinfo{volume}{6}}, \bibinfo{pages}{256}
  (\bibinfo{year}{2007}).

\bibitem[{\citenamefont{Roy et~al.}(2012)\citenamefont{Roy, Gupta, and
  Garg}}]{Roy12}
\bibinfo{author}{\bibfnamefont{A.}~\bibnamefont{Roy}},
  \bibinfo{author}{\bibfnamefont{R.}~\bibnamefont{Gupta}}, \bibnamefont{and}
  \bibinfo{author}{\bibfnamefont{A.}~\bibnamefont{Garg}},
  \bibinfo{journal}{Adv. in Cond. Matt. Phys.} \textbf{\bibinfo{volume}{2012}},
  \bibinfo{pages}{926290} (\bibinfo{year}{2012}).

\bibitem[{\citenamefont{Kruglyak et~al.}(2010)\citenamefont{Kruglyak,
  Demokritov, and Grundler}}]{Kruglyak10}
\bibinfo{author}{\bibfnamefont{V.~V.} \bibnamefont{Kruglyak}},
  \bibinfo{author}{\bibfnamefont{S.~O.} \bibnamefont{Demokritov}},
  \bibnamefont{and} \bibinfo{author}{\bibfnamefont{D.}~\bibnamefont{Grundler}},
  \bibinfo{journal}{J. Phys. D: Appl. Physics} \textbf{\bibinfo{volume}{43}},
  \bibinfo{pages}{264001} (\bibinfo{year}{2010}).

\bibitem[{\citenamefont{Nan et~al.}(2008)\citenamefont{Nan, Bichurin, Dong,
  Viehland, and Srinivasan}}]{Nan08}
\bibinfo{author}{\bibfnamefont{C.}~\bibnamefont{Nan}},
  \bibinfo{author}{\bibfnamefont{M.}~\bibnamefont{Bichurin}},
  \bibinfo{author}{\bibfnamefont{S.}~\bibnamefont{Dong}},
  \bibinfo{author}{\bibfnamefont{D.}~\bibnamefont{Viehland}}, \bibnamefont{and}
  \bibinfo{author}{\bibfnamefont{G.}~\bibnamefont{Srinivasan}},
  \bibinfo{journal}{J. Appl. Phys.} \textbf{\bibinfo{volume}{103}},
  \bibinfo{pages}{031101} (\bibinfo{year}{2008}).

\bibitem[{\citenamefont{Pimenov et~al.}(2006)\citenamefont{Pimenov, Mukhin,
  Ivanov, Travkin, Balbashov, and Loidl}}]{Pimenov06}
\bibinfo{author}{\bibfnamefont{A.}~\bibnamefont{Pimenov}},
  \bibinfo{author}{\bibfnamefont{A.}~\bibnamefont{Mukhin}},
  \bibinfo{author}{\bibfnamefont{V.}~\bibnamefont{Ivanov}},
  \bibinfo{author}{\bibfnamefont{V.}~\bibnamefont{Travkin}},
  \bibinfo{author}{\bibfnamefont{A.}~\bibnamefont{Balbashov}},
  \bibnamefont{and} \bibinfo{author}{\bibfnamefont{A.}~\bibnamefont{Loidl}},
  \bibinfo{journal}{Nat. Phys.} \textbf{\bibinfo{volume}{2}},
  \bibinfo{pages}{97} (\bibinfo{year}{2006}).

\bibitem[{\citenamefont{Vald\'es~Aguilar
  et~al.}(2007)\citenamefont{Vald\'es~Aguilar, Sushkov, Zhang, Choi, Cheong,
  and Drew}}]{ValdesAguilar07}
\bibinfo{author}{\bibfnamefont{R.}~\bibnamefont{Vald\'es~Aguilar}},
  \bibinfo{author}{\bibfnamefont{A.~B.} \bibnamefont{Sushkov}},
  \bibinfo{author}{\bibfnamefont{C.~L.} \bibnamefont{Zhang}},
  \bibinfo{author}{\bibfnamefont{Y.~J.} \bibnamefont{Choi}},
  \bibinfo{author}{\bibfnamefont{S.-W.} \bibnamefont{Cheong}},
  \bibnamefont{and} \bibinfo{author}{\bibfnamefont{H.~D.} \bibnamefont{Drew}},
  \bibinfo{journal}{Phys. Rev. B} \textbf{\bibinfo{volume}{76}},
  \bibinfo{pages}{060404} (\bibinfo{year}{2007}).

\bibitem[{\citenamefont{Vald\'es~Aguilar
  et~al.}(2009)\citenamefont{Vald\'es~Aguilar, Mostovoy, Sushkov, Zhang, Choi,
  Cheong, and Drew}}]{ValdesAguilar09}
\bibinfo{author}{\bibfnamefont{R.}~\bibnamefont{Vald\'es~Aguilar}},
  \bibinfo{author}{\bibfnamefont{M.}~\bibnamefont{Mostovoy}},
  \bibinfo{author}{\bibfnamefont{A.~B.} \bibnamefont{Sushkov}},
  \bibinfo{author}{\bibfnamefont{C.~L.} \bibnamefont{Zhang}},
  \bibinfo{author}{\bibfnamefont{Y.~J.} \bibnamefont{Choi}},
  \bibinfo{author}{\bibfnamefont{S.-W.} \bibnamefont{Cheong}},
  \bibnamefont{and} \bibinfo{author}{\bibfnamefont{H.~D.} \bibnamefont{Drew}},
  \bibinfo{journal}{Phys. Rev. Lett.} \textbf{\bibinfo{volume}{102}},
  \bibinfo{pages}{047203} (\bibinfo{year}{2009}).

\bibitem[{\citenamefont{Takahashi et~al.}(2012)\citenamefont{Takahashi,
  Shimano, Kaneko, Murakawa, and Tokura}}]{Takahashi12}
\bibinfo{author}{\bibfnamefont{Y.}~\bibnamefont{Takahashi}},
  \bibinfo{author}{\bibfnamefont{R.}~\bibnamefont{Shimano}},
  \bibinfo{author}{\bibfnamefont{Y.}~\bibnamefont{Kaneko}},
  \bibinfo{author}{\bibfnamefont{H.}~\bibnamefont{Murakawa}}, \bibnamefont{and}
  \bibinfo{author}{\bibfnamefont{Y.}~\bibnamefont{Tokura}},
  \bibinfo{journal}{Nat. Phys.} \textbf{\bibinfo{volume}{8}},
  \bibinfo{pages}{121} (\bibinfo{year}{2012}).

\bibitem[{\citenamefont{Stenberg and de~Sousa}(2012)}]{Stenberg12}
\bibinfo{author}{\bibfnamefont{M.~P.~V.} \bibnamefont{Stenberg}}
  \bibnamefont{and} \bibinfo{author}{\bibfnamefont{R.}~\bibnamefont{de~Sousa}},
  \bibinfo{journal}{Phys. Rev. B} \textbf{\bibinfo{volume}{85}},
  \bibinfo{pages}{104412} (\bibinfo{year}{2012}), \bibinfo{note}{ibid.
  \textbf{80}, 094419 (2009)}.

\bibitem[{\citenamefont{Mochizuki et~al.}(2010)\citenamefont{Mochizuki,
  Furukawa, and Nagaosa}}]{Mochizuki10}
\bibinfo{author}{\bibfnamefont{M.}~\bibnamefont{Mochizuki}},
  \bibinfo{author}{\bibfnamefont{N.}~\bibnamefont{Furukawa}}, \bibnamefont{and}
  \bibinfo{author}{\bibfnamefont{N.}~\bibnamefont{Nagaosa}},
  \bibinfo{journal}{Phys. Rev. Lett.} \textbf{\bibinfo{volume}{104}},
  \bibinfo{pages}{177206} (\bibinfo{year}{2010}).

\bibitem[{\citenamefont{Khomskii}(2009)}]{Khomskii09}
\bibinfo{author}{\bibfnamefont{D.}~\bibnamefont{Khomskii}},
  \bibinfo{journal}{Phys.} \textbf{\bibinfo{volume}{2}}, \bibinfo{pages}{1}
  (\bibinfo{year}{2009}).

\bibitem[{\citenamefont{Sushkov et~al.}(2007)\citenamefont{Sushkov, Aguilar,
  Park, Cheong, and Drew}}]{Sushkov07}
\bibinfo{author}{\bibfnamefont{A.}~\bibnamefont{Sushkov}},
  \bibinfo{author}{\bibfnamefont{R.}~\bibnamefont{Aguilar}},
  \bibinfo{author}{\bibfnamefont{S.}~\bibnamefont{Park}},
  \bibinfo{author}{\bibfnamefont{S.}~\bibnamefont{Cheong}}, \bibnamefont{and}
  \bibinfo{author}{\bibfnamefont{H.}~\bibnamefont{Drew}},
  \bibinfo{journal}{Phys. Rev. Lett.} \textbf{\bibinfo{volume}{98}},
  \bibinfo{pages}{27202} (\bibinfo{year}{2007}).

\bibitem[{\citenamefont{Sushkov et~al.}(2008)\citenamefont{Sushkov, Mostovoy,
  Aguilar, Cheong, and Drew}}]{Sushkov08}
\bibinfo{author}{\bibfnamefont{A.~B.} \bibnamefont{Sushkov}},
  \bibinfo{author}{\bibfnamefont{M.}~\bibnamefont{Mostovoy}},
  \bibinfo{author}{\bibfnamefont{R.~V.} \bibnamefont{Aguilar}},
  \bibinfo{author}{\bibfnamefont{S.-W.} \bibnamefont{Cheong}},
  \bibnamefont{and} \bibinfo{author}{\bibfnamefont{H.~D.} \bibnamefont{Drew}},
  \bibinfo{journal}{J. Phys.: Cond. Matt.} \textbf{\bibinfo{volume}{20}},
  \bibinfo{pages}{434210} (\bibinfo{year}{2008}).

\bibitem[{\citenamefont{Pimenov et~al.}(2008)\citenamefont{Pimenov, Shuvaev,
  Mukhin, and Loidl}}]{Pimenov08}
\bibinfo{author}{\bibfnamefont{A.}~\bibnamefont{Pimenov}},
  \bibinfo{author}{\bibfnamefont{A.~M.} \bibnamefont{Shuvaev}},
  \bibinfo{author}{\bibfnamefont{A.~A.} \bibnamefont{Mukhin}},
  \bibnamefont{and} \bibinfo{author}{\bibfnamefont{A.}~\bibnamefont{Loidl}},
  \bibinfo{journal}{J. Phys.: Cond. Matt.} \textbf{\bibinfo{volume}{20}},
  \bibinfo{pages}{434209} (\bibinfo{year}{2008}).

\bibitem[{\citenamefont{Kida et~al.}(2009)\citenamefont{Kida, Takahashi, Lee,
  Shimano, Yamasaki, Kaneko, Miyahara, Furukawa, Arima, and Tokura}}]{Kida09}
\bibinfo{author}{\bibfnamefont{N.}~\bibnamefont{Kida}},
  \bibinfo{author}{\bibfnamefont{Y.}~\bibnamefont{Takahashi}},
  \bibinfo{author}{\bibfnamefont{J.}~\bibnamefont{Lee}},
  \bibinfo{author}{\bibfnamefont{R.}~\bibnamefont{Shimano}},
  \bibinfo{author}{\bibfnamefont{Y.}~\bibnamefont{Yamasaki}},
  \bibinfo{author}{\bibfnamefont{Y.}~\bibnamefont{Kaneko}},
  \bibinfo{author}{\bibfnamefont{S.}~\bibnamefont{Miyahara}},
  \bibinfo{author}{\bibfnamefont{N.}~\bibnamefont{Furukawa}},
  \bibinfo{author}{\bibfnamefont{T.}~\bibnamefont{Arima}}, \bibnamefont{and}
  \bibinfo{author}{\bibfnamefont{Y.}~\bibnamefont{Tokura}},
  \bibinfo{journal}{J. Opt. Soc. Am. B} \textbf{\bibinfo{volume}{26}},
  \bibinfo{pages}{A35} (\bibinfo{year}{2009}).

\bibitem[{\citenamefont{Seki et~al.}(2010)\citenamefont{Seki, Kida, Kumakura,
  Shimano, and Tokura}}]{Seki10}
\bibinfo{author}{\bibfnamefont{S.}~\bibnamefont{Seki}},
  \bibinfo{author}{\bibfnamefont{N.}~\bibnamefont{Kida}},
  \bibinfo{author}{\bibfnamefont{S.}~\bibnamefont{Kumakura}},
  \bibinfo{author}{\bibfnamefont{R.}~\bibnamefont{Shimano}}, \bibnamefont{and}
  \bibinfo{author}{\bibfnamefont{Y.}~\bibnamefont{Tokura}},
  \bibinfo{journal}{Phys. Rev. Lett.} \textbf{\bibinfo{volume}{105}},
  \bibinfo{pages}{097207} (\bibinfo{year}{2010}).

\bibitem[{\citenamefont{K{\'e}zsm{\'a}rki
  et~al.}(2011)\citenamefont{K{\'e}zsm{\'a}rki, Kida, Murakawa, Bord{\'a}cs,
  Onose, and Tokura}}]{Kezsmarki11}
\bibinfo{author}{\bibfnamefont{I.}~\bibnamefont{K{\'e}zsm{\'a}rki}},
  \bibinfo{author}{\bibfnamefont{N.}~\bibnamefont{Kida}},
  \bibinfo{author}{\bibfnamefont{H.}~\bibnamefont{Murakawa}},
  \bibinfo{author}{\bibfnamefont{S.}~\bibnamefont{Bord{\'a}cs}},
  \bibinfo{author}{\bibfnamefont{Y.}~\bibnamefont{Onose}}, \bibnamefont{and}
  \bibinfo{author}{\bibfnamefont{Y.}~\bibnamefont{Tokura}},
  \bibinfo{journal}{Phys. Rev. Lett.} \textbf{\bibinfo{volume}{106}},
  \bibinfo{pages}{57403} (\bibinfo{year}{2011}).

\bibitem[{\citenamefont{Shuvaev et~al.}(2011)\citenamefont{Shuvaev, Mukhin, and
  Pimenov}}]{Shuvaev11}
\bibinfo{author}{\bibfnamefont{A.~M.} \bibnamefont{Shuvaev}},
  \bibinfo{author}{\bibfnamefont{A.~A.} \bibnamefont{Mukhin}},
  \bibnamefont{and} \bibinfo{author}{\bibfnamefont{A.}~\bibnamefont{Pimenov}},
  \bibinfo{journal}{J. Phys.: Cond. Matt.} \textbf{\bibinfo{volume}{23}},
  \bibinfo{pages}{113201} (\bibinfo{year}{2011}).

\bibitem[{\citenamefont{Cazayous et~al.}(2008)\citenamefont{Cazayous, Gallais,
  Sacuto, De~Sousa, Lebeugle, and Colson}}]{Cazayous08}
\bibinfo{author}{\bibfnamefont{M.}~\bibnamefont{Cazayous}},
  \bibinfo{author}{\bibfnamefont{Y.}~\bibnamefont{Gallais}},
  \bibinfo{author}{\bibfnamefont{A.}~\bibnamefont{Sacuto}},
  \bibinfo{author}{\bibfnamefont{R.}~\bibnamefont{De~Sousa}},
  \bibinfo{author}{\bibfnamefont{D.}~\bibnamefont{Lebeugle}}, \bibnamefont{and}
  \bibinfo{author}{\bibfnamefont{D.}~\bibnamefont{Colson}},
  \bibinfo{journal}{Phys. Rev. Lett.} \textbf{\bibinfo{volume}{101}},
  \bibinfo{pages}{37601} (\bibinfo{year}{2008}).

\bibitem[{\citenamefont{Talbayev et~al.}(2011)\citenamefont{Talbayev, Trugman,
  Lee, Yi, Cheong, and Taylor}}]{Talbayev11}
\bibinfo{author}{\bibfnamefont{D.}~\bibnamefont{Talbayev}},
  \bibinfo{author}{\bibfnamefont{S.}~\bibnamefont{Trugman}},
  \bibinfo{author}{\bibfnamefont{S.}~\bibnamefont{Lee}},
  \bibinfo{author}{\bibfnamefont{H.}~\bibnamefont{Yi}},
  \bibinfo{author}{\bibfnamefont{S.}~\bibnamefont{Cheong}}, \bibnamefont{and}
  \bibinfo{author}{\bibfnamefont{A.}~\bibnamefont{Taylor}},
  \bibinfo{journal}{Phys. Rev. B} \textbf{\bibinfo{volume}{83}},
  \bibinfo{pages}{094403} (\bibinfo{year}{2011}).

\bibitem[{\citenamefont{Komandin et~al.}(2010)\citenamefont{Komandin,
  Torgashev, Volkov, Porodinkov, Spektor, and Bush}}]{Komandin10}
\bibinfo{author}{\bibfnamefont{G.}~\bibnamefont{Komandin}},
  \bibinfo{author}{\bibfnamefont{V.}~\bibnamefont{Torgashev}},
  \bibinfo{author}{\bibfnamefont{A.}~\bibnamefont{Volkov}},
  \bibinfo{author}{\bibfnamefont{O.}~\bibnamefont{Porodinkov}},
  \bibinfo{author}{\bibfnamefont{I.}~\bibnamefont{Spektor}}, \bibnamefont{and}
  \bibinfo{author}{\bibfnamefont{A.}~\bibnamefont{Bush}},
  \bibinfo{journal}{Phys. Solid State} \textbf{\bibinfo{volume}{52}},
  \bibinfo{pages}{734} (\bibinfo{year}{2010}).

\bibitem[{\citenamefont{Pailh\`es et~al.}(2009)\citenamefont{Pailh\`es,
  Fabr\`eges, R\'egnault, Pinsard-Godart, Mirebeau, Moussa, Hennion, and
  Petit}}]{Pailhes09}
\bibinfo{author}{\bibfnamefont{S.}~\bibnamefont{Pailh\`es}},
  \bibinfo{author}{\bibfnamefont{X.}~\bibnamefont{Fabr\`eges}},
  \bibinfo{author}{\bibfnamefont{L.~P.} \bibnamefont{R\'egnault}},
  \bibinfo{author}{\bibfnamefont{L.}~\bibnamefont{Pinsard-Godart}},
  \bibinfo{author}{\bibfnamefont{I.}~\bibnamefont{Mirebeau}},
  \bibinfo{author}{\bibfnamefont{F.}~\bibnamefont{Moussa}},
  \bibinfo{author}{\bibfnamefont{M.}~\bibnamefont{Hennion}}, \bibnamefont{and}
  \bibinfo{author}{\bibfnamefont{S.}~\bibnamefont{Petit}},
  \bibinfo{journal}{Phys. Rev. B} \textbf{\bibinfo{volume}{79}},
  \bibinfo{pages}{134409} (\bibinfo{year}{2009}).

\bibitem[{\citenamefont{Kadlec et~al.}(2011)\citenamefont{Kadlec, Goian,
  Rushchanskii, Ku\ifmmode~\check{z}\else \v{z}\fi{}el, Le\ifmmode
  \check{z}\else \v{z}\fi{}ai\ifmmode~\acute{c}\else \'{c}\fi{}, Kohn, Pisarev,
  and Kamba}}]{Kadlec11}
\bibinfo{author}{\bibfnamefont{C.}~\bibnamefont{Kadlec}},
  \bibinfo{author}{\bibfnamefont{V.}~\bibnamefont{Goian}},
  \bibinfo{author}{\bibfnamefont{K.~Z.} \bibnamefont{Rushchanskii}},
  \bibinfo{author}{\bibfnamefont{P.}~\bibnamefont{Ku\ifmmode~\check{z}\else
  \v{z}\fi{}el}}, \bibinfo{author}{\bibfnamefont{M.}~\bibnamefont{Le\ifmmode
  \check{z}\else \v{z}\fi{}ai\ifmmode~\acute{c}\else \'{c}\fi{}}},
  \bibinfo{author}{\bibfnamefont{K.}~\bibnamefont{Kohn}},
  \bibinfo{author}{\bibfnamefont{R.~V.} \bibnamefont{Pisarev}},
  \bibnamefont{and} \bibinfo{author}{\bibfnamefont{S.}~\bibnamefont{Kamba}},
  \bibinfo{journal}{Phys. Rev. B} \textbf{\bibinfo{volume}{84}},
  \bibinfo{pages}{174120} (\bibinfo{year}{2011}).

\bibitem[{\citenamefont{Machala et~al.}(2011)\citenamefont{Machala, Tu\v{c}ek,
  and Zbo\v{r}il}}]{Machala11}
\bibinfo{author}{\bibfnamefont{L.}~\bibnamefont{Machala}},
  \bibinfo{author}{\bibfnamefont{J.}~\bibnamefont{Tu\v{c}ek}},
  \bibnamefont{and}
  \bibinfo{author}{\bibfnamefont{R.}~\bibnamefont{Zbo\v{r}il}},
  \bibinfo{journal}{Chem. Mater.} \textbf{\bibinfo{volume}{23}},
  \bibinfo{pages}{3255} (\bibinfo{year}{2011}).

\bibitem[{\citenamefont{Namai et~al.}(2008)\citenamefont{Namai, Sakurai,
  Nakajima, Suemoto, Matsumoto, Goto, Sasaki, and Ohkoshi}}]{Namai08}
\bibinfo{author}{\bibfnamefont{A.}~\bibnamefont{Namai}},
  \bibinfo{author}{\bibfnamefont{S.}~\bibnamefont{Sakurai}},
  \bibinfo{author}{\bibfnamefont{M.}~\bibnamefont{Nakajima}},
  \bibinfo{author}{\bibfnamefont{T.}~\bibnamefont{Suemoto}},
  \bibinfo{author}{\bibfnamefont{K.}~\bibnamefont{Matsumoto}},
  \bibinfo{author}{\bibfnamefont{M.}~\bibnamefont{Goto}},
  \bibinfo{author}{\bibfnamefont{S.}~\bibnamefont{Sasaki}}, \bibnamefont{and}
  \bibinfo{author}{\bibfnamefont{S.}~\bibnamefont{Ohkoshi}},
  \bibinfo{journal}{J. Am. Chem. Soc.} \textbf{\bibinfo{volume}{131}},
  \bibinfo{pages}{1170} (\bibinfo{year}{2008}).

\bibitem[{\citenamefont{Namai et~al.}(2012)\citenamefont{Namai, Yoshikiyo,
  Yamada, Sakurai, Goto, Yoshida, Miyazaki, Nakajima, Suemoto, and
  Tokoro}}]{Namai12}
\bibinfo{author}{\bibfnamefont{A.}~\bibnamefont{Namai}},
  \bibinfo{author}{\bibfnamefont{M.}~\bibnamefont{Yoshikiyo}},
  \bibinfo{author}{\bibfnamefont{K.}~\bibnamefont{Yamada}},
  \bibinfo{author}{\bibfnamefont{S.}~\bibnamefont{Sakurai}},
  \bibinfo{author}{\bibfnamefont{T.}~\bibnamefont{Goto}},
  \bibinfo{author}{\bibfnamefont{T.}~\bibnamefont{Yoshida}},
  \bibinfo{author}{\bibfnamefont{T.}~\bibnamefont{Miyazaki}},
  \bibinfo{author}{\bibfnamefont{M.}~\bibnamefont{Nakajima}},
  \bibinfo{author}{\bibfnamefont{T.}~\bibnamefont{Suemoto}}, \bibnamefont{and}
  \bibinfo{author}{\bibfnamefont{H.}~\bibnamefont{Tokoro}},
  \bibinfo{journal}{Nat. Commun.} \textbf{\bibinfo{volume}{3}},
  \bibinfo{pages}{1035} (\bibinfo{year}{2012}).

\bibitem[{\citenamefont{Ohkoshi et~al.}(2007)\citenamefont{Ohkoshi, Kuroki,
  Sakurai, Matsumoto, Sato, and Sasaki}}]{Ohkoshi07}
\bibinfo{author}{\bibfnamefont{S.}~\bibnamefont{Ohkoshi}},
  \bibinfo{author}{\bibfnamefont{S.}~\bibnamefont{Kuroki}},
  \bibinfo{author}{\bibfnamefont{S.}~\bibnamefont{Sakurai}},
  \bibinfo{author}{\bibfnamefont{K.}~\bibnamefont{Matsumoto}},
  \bibinfo{author}{\bibfnamefont{K.}~\bibnamefont{Sato}}, \bibnamefont{and}
  \bibinfo{author}{\bibfnamefont{S.}~\bibnamefont{Sasaki}},
  \bibinfo{journal}{Angew. Chem. Int. Ed.} \textbf{\bibinfo{volume}{46}},
  \bibinfo{pages}{8392} (\bibinfo{year}{2007}).

\bibitem[{\citenamefont{Tu\v{c}ek et~al.}(2010)\citenamefont{Tu\v{c}ek,
  Zbo\v{r}il, Namai, and Ohkoshi}}]{Tucek10}
\bibinfo{author}{\bibfnamefont{J.}~\bibnamefont{Tu\v{c}ek}},
  \bibinfo{author}{\bibfnamefont{R.}~\bibnamefont{Zbo\v{r}il}},
  \bibinfo{author}{\bibfnamefont{A.}~\bibnamefont{Namai}}, \bibnamefont{and}
  \bibinfo{author}{\bibfnamefont{S.}~\bibnamefont{Ohkoshi}},
  \bibinfo{journal}{Chem. Mater.} \textbf{\bibinfo{volume}{22}},
  \bibinfo{pages}{6483} (\bibinfo{year}{2010}).

\bibitem[{\citenamefont{Gich et~al.}(2010)\citenamefont{Gich, Gazquez, Roig,
  Crespi, Fontcuberta, Idrobo, Pennycook, Varela, Skumryev, and
  Varela}}]{Gich10}
\bibinfo{author}{\bibfnamefont{M.}~\bibnamefont{Gich}},
  \bibinfo{author}{\bibfnamefont{J.}~\bibnamefont{Gazquez}},
  \bibinfo{author}{\bibfnamefont{A.}~\bibnamefont{Roig}},
  \bibinfo{author}{\bibfnamefont{A.}~\bibnamefont{Crespi}},
  \bibinfo{author}{\bibfnamefont{J.}~\bibnamefont{Fontcuberta}},
  \bibinfo{author}{\bibfnamefont{J.~C.} \bibnamefont{Idrobo}},
  \bibinfo{author}{\bibfnamefont{S.~J.} \bibnamefont{Pennycook}},
  \bibinfo{author}{\bibfnamefont{M.}~\bibnamefont{Varela}},
  \bibinfo{author}{\bibfnamefont{V.}~\bibnamefont{Skumryev}}, \bibnamefont{and}
  \bibinfo{author}{\bibfnamefont{M.}~\bibnamefont{Varela}},
  \bibinfo{journal}{Appl. Phys. Lett.} \textbf{\bibinfo{volume}{96}},
  \bibinfo{pages}{112508} (\bibinfo{year}{2010}).

\bibitem[{\citenamefont{Ding et~al.}(2007)\citenamefont{Ding, Morber, Snyder,
  and Wang}}]{Ding07}
\bibinfo{author}{\bibfnamefont{Y.}~\bibnamefont{Ding}},
  \bibinfo{author}{\bibfnamefont{J.}~\bibnamefont{Morber}},
  \bibinfo{author}{\bibfnamefont{R.}~\bibnamefont{Snyder}}, \bibnamefont{and}
  \bibinfo{author}{\bibfnamefont{Z.}~\bibnamefont{Wang}},
  \bibinfo{journal}{Adv. Funct. Mater.} \textbf{\bibinfo{volume}{17}},
  \bibinfo{pages}{1172} (\bibinfo{year}{2007}).

\bibitem[{\citenamefont{Jin et~al.}(2004)\citenamefont{Jin, Ohkoshi, and
  Hashimoto}}]{Jin04}
\bibinfo{author}{\bibfnamefont{J.}~\bibnamefont{Jin}},
  \bibinfo{author}{\bibfnamefont{S.}~\bibnamefont{Ohkoshi}}, \bibnamefont{and}
  \bibinfo{author}{\bibfnamefont{K.}~\bibnamefont{Hashimoto}},
  \bibinfo{journal}{Adv. Mater.} \textbf{\bibinfo{volume}{16}},
  \bibinfo{pages}{48} (\bibinfo{year}{2004}).

\bibitem[{\citenamefont{Sakurai et~al.}(2005)\citenamefont{Sakurai, Jin,
  Hashimoto, and Ohkoshi}}]{Sakurai05}
\bibinfo{author}{\bibfnamefont{S.}~\bibnamefont{Sakurai}},
  \bibinfo{author}{\bibfnamefont{J.}~\bibnamefont{Jin}},
  \bibinfo{author}{\bibfnamefont{K.}~\bibnamefont{Hashimoto}},
  \bibnamefont{and} \bibinfo{author}{\bibfnamefont{S.}~\bibnamefont{Ohkoshi}},
  \bibinfo{journal}{J. Phys. Soc. Japan} \textbf{\bibinfo{volume}{74}},
  \bibinfo{pages}{1946} (\bibinfo{year}{2005}).

\bibitem[{\citenamefont{Tu\v{c}ek et~al.}(2011)\citenamefont{Tu\v{c}ek,
  Ohkoshi, and Zbo\v{r}il}}]{Tucek11}
\bibinfo{author}{\bibfnamefont{J.}~\bibnamefont{Tu\v{c}ek}},
  \bibinfo{author}{\bibfnamefont{S.}~\bibnamefont{Ohkoshi}}, \bibnamefont{and}
  \bibinfo{author}{\bibfnamefont{R.}~\bibnamefont{Zbo\v{r}il}},
  \bibinfo{journal}{Appl. Phys. Lett.} \textbf{\bibinfo{volume}{99}},
  \bibinfo{pages}{253108} (\bibinfo{year}{2011}).

\bibitem[{\citenamefont{Tronc et~al.}(1998)\citenamefont{Tronc, Chan\'eac, and
  Jolivet}}]{Tronc98}
\bibinfo{author}{\bibfnamefont{E.}~\bibnamefont{Tronc}},
  \bibinfo{author}{\bibfnamefont{C.}~\bibnamefont{Chan\'eac}},
  \bibnamefont{and} \bibinfo{author}{\bibfnamefont{J.}~\bibnamefont{Jolivet}},
  \bibinfo{journal}{Journal of Solid State Chemistry}
  \textbf{\bibinfo{volume}{139}}, \bibinfo{pages}{93 } (\bibinfo{year}{1998}).

\bibitem[{\citenamefont{Gich et~al.}(2006{\natexlab{a}})\citenamefont{Gich,
  Frontera, Roig, Taboada, Molins, Rechenberg, Ardisson, Macedo, Ritter, Hardy
  et~al.}}]{Gich06}
\bibinfo{author}{\bibfnamefont{M.}~\bibnamefont{Gich}},
  \bibinfo{author}{\bibfnamefont{C.}~\bibnamefont{Frontera}},
  \bibinfo{author}{\bibfnamefont{A.}~\bibnamefont{Roig}},
  \bibinfo{author}{\bibfnamefont{E.}~\bibnamefont{Taboada}},
  \bibinfo{author}{\bibfnamefont{E.}~\bibnamefont{Molins}},
  \bibinfo{author}{\bibfnamefont{H.~R.} \bibnamefont{Rechenberg}},
  \bibinfo{author}{\bibfnamefont{J.~D.} \bibnamefont{Ardisson}},
  \bibinfo{author}{\bibfnamefont{W.~A.~A.} \bibnamefont{Macedo}},
  \bibinfo{author}{\bibfnamefont{C.}~\bibnamefont{Ritter}},
  \bibinfo{author}{\bibfnamefont{V.}~\bibnamefont{Hardy}},
  \bibnamefont{et~al.}, \bibinfo{journal}{Chem. Mater.}
  \textbf{\bibinfo{volume}{18}}, \bibinfo{pages}{3889}
  (\bibinfo{year}{2006}{\natexlab{a}}).

\bibitem[{\citenamefont{Gich et~al.}(2006{\natexlab{b}})\citenamefont{Gich,
  Frontera, Roig, Fontcuberta, Molins, Bellido, Simon, and Fleta}}]{Gich06a}
\bibinfo{author}{\bibfnamefont{M.}~\bibnamefont{Gich}},
  \bibinfo{author}{\bibfnamefont{C.}~\bibnamefont{Frontera}},
  \bibinfo{author}{\bibfnamefont{A.}~\bibnamefont{Roig}},
  \bibinfo{author}{\bibfnamefont{J.}~\bibnamefont{Fontcuberta}},
  \bibinfo{author}{\bibfnamefont{E.}~\bibnamefont{Molins}},
  \bibinfo{author}{\bibfnamefont{N.}~\bibnamefont{Bellido}},
  \bibinfo{author}{\bibfnamefont{C.}~\bibnamefont{Simon}}, \bibnamefont{and}
  \bibinfo{author}{\bibfnamefont{C.}~\bibnamefont{Fleta}},
  \bibinfo{journal}{Nanotechnology} \textbf{\bibinfo{volume}{17}},
  \bibinfo{pages}{687} (\bibinfo{year}{2006}{\natexlab{b}}).

\bibitem[{\citenamefont{Tseng et~al.}(2009)\citenamefont{Tseng, Souza-Neto,
  Haskel, Gich, Frontera, Roig, van Veenendaal, and Nogu\'es}}]{Tseng09}
\bibinfo{author}{\bibfnamefont{Y.-C.} \bibnamefont{Tseng}},
  \bibinfo{author}{\bibfnamefont{N.~M.} \bibnamefont{Souza-Neto}},
  \bibinfo{author}{\bibfnamefont{D.}~\bibnamefont{Haskel}},
  \bibinfo{author}{\bibfnamefont{M.}~\bibnamefont{Gich}},
  \bibinfo{author}{\bibfnamefont{C.}~\bibnamefont{Frontera}},
  \bibinfo{author}{\bibfnamefont{A.}~\bibnamefont{Roig}},
  \bibinfo{author}{\bibfnamefont{M.}~\bibnamefont{van Veenendaal}},
  \bibnamefont{and} \bibinfo{author}{\bibfnamefont{J.}~\bibnamefont{Nogu\'es}},
  \bibinfo{journal}{Phys. Rev. B} \textbf{\bibinfo{volume}{79}},
  \bibinfo{pages}{094404} (\bibinfo{year}{2009}).

\bibitem[{\citenamefont{Kant et~al.}(2012)\citenamefont{Kant, Schmidt, Wang,
  Mayr, Tsurkan, Deisenhofer, and Loidl}}]{Kant12}
\bibinfo{author}{\bibfnamefont{C.}~\bibnamefont{Kant}},
  \bibinfo{author}{\bibfnamefont{M.}~\bibnamefont{Schmidt}},
  \bibinfo{author}{\bibfnamefont{Z.}~\bibnamefont{Wang}},
  \bibinfo{author}{\bibfnamefont{F.}~\bibnamefont{Mayr}},
  \bibinfo{author}{\bibfnamefont{V.}~\bibnamefont{Tsurkan}},
  \bibinfo{author}{\bibfnamefont{J.}~\bibnamefont{Deisenhofer}},
  \bibnamefont{and} \bibinfo{author}{\bibfnamefont{A.}~\bibnamefont{Loidl}},
  \bibinfo{journal}{Phys. Rev. Lett.} \textbf{\bibinfo{volume}{108}},
  \bibinfo{pages}{177203} (\bibinfo{year}{2012}).

\bibitem[{\citenamefont{Gich et~al.}(2005)\citenamefont{Gich, Roig, Frontera,
  Molins, Sort, Popovici, Chouteau, y~Marero, and Nogu\'{e}s}}]{Gich05}
\bibinfo{author}{\bibfnamefont{M.}~\bibnamefont{Gich}},
  \bibinfo{author}{\bibfnamefont{A.}~\bibnamefont{Roig}},
  \bibinfo{author}{\bibfnamefont{C.}~\bibnamefont{Frontera}},
  \bibinfo{author}{\bibfnamefont{E.}~\bibnamefont{Molins}},
  \bibinfo{author}{\bibfnamefont{J.}~\bibnamefont{Sort}},
  \bibinfo{author}{\bibfnamefont{M.}~\bibnamefont{Popovici}},
  \bibinfo{author}{\bibfnamefont{G.}~\bibnamefont{Chouteau}},
  \bibinfo{author}{\bibfnamefont{D.~M.} \bibnamefont{y~Marero}},
  \bibnamefont{and}
  \bibinfo{author}{\bibfnamefont{J.}~\bibnamefont{Nogu\'{e}s}},
  \bibinfo{journal}{J. Appl. Phys.} \textbf{\bibinfo{volume}{98}},
  \bibinfo{pages}{044307} (\bibinfo{year}{2005}).

\bibitem[{\citenamefont{Schrettle et~al.}(2009)\citenamefont{Schrettle,
  Lunkenheimer, Hemberger, Ivanov, Mukhin, Balbashov, and Loidl}}]{Schrettle09}
\bibinfo{author}{\bibfnamefont{F.}~\bibnamefont{Schrettle}},
  \bibinfo{author}{\bibfnamefont{P.}~\bibnamefont{Lunkenheimer}},
  \bibinfo{author}{\bibfnamefont{J.}~\bibnamefont{Hemberger}},
  \bibinfo{author}{\bibfnamefont{V.~Y.} \bibnamefont{Ivanov}},
  \bibinfo{author}{\bibfnamefont{A.~A.} \bibnamefont{Mukhin}},
  \bibinfo{author}{\bibfnamefont{A.~M.} \bibnamefont{Balbashov}},
  \bibnamefont{and} \bibinfo{author}{\bibfnamefont{A.}~\bibnamefont{Loidl}},
  \bibinfo{journal}{Phys. Rev. Lett.} \textbf{\bibinfo{volume}{102}},
  \bibinfo{pages}{207208} (\bibinfo{year}{2009}).

\bibitem[{\citenamefont{Pyatakov et~al.}(2011)\citenamefont{Pyatakov, Sechin,
  Sergeev, Nikolaev, Nikolaeva, Logginov, and Zvezdin}}]{Pyatakov11}
\bibinfo{author}{\bibfnamefont{A.}~\bibnamefont{Pyatakov}},
  \bibinfo{author}{\bibfnamefont{D.}~\bibnamefont{Sechin}},
  \bibinfo{author}{\bibfnamefont{A.}~\bibnamefont{Sergeev}},
  \bibinfo{author}{\bibfnamefont{A.}~\bibnamefont{Nikolaev}},
  \bibinfo{author}{\bibfnamefont{E.}~\bibnamefont{Nikolaeva}},
  \bibinfo{author}{\bibfnamefont{A.}~\bibnamefont{Logginov}}, \bibnamefont{and}
  \bibinfo{author}{\bibfnamefont{A.}~\bibnamefont{Zvezdin}},
  \bibinfo{journal}{Europhys. Lett.} \textbf{\bibinfo{volume}{93}},
  \bibinfo{pages}{17001} (\bibinfo{year}{2011}).

\bibitem[{\citenamefont{Gich et~al.}(in preparation)\citenamefont{Gich, Fina,
  Morelli, S\'anchez, Alexe, Fontcuberta, and Roig}}]{Gich-prep}
\bibinfo{author}{\bibfnamefont{M.}~\bibnamefont{Gich}},
  \bibinfo{author}{\bibfnamefont{I.}~\bibnamefont{Fina}},
  \bibinfo{author}{\bibfnamefont{A.}~\bibnamefont{Morelli}},
  \bibinfo{author}{\bibfnamefont{F.}~\bibnamefont{S\'anchez}},
  \bibinfo{author}{\bibfnamefont{M.}~\bibnamefont{Alexe}},
  \bibinfo{author}{\bibfnamefont{J.}~\bibnamefont{Fontcuberta}},
  \bibnamefont{and} \bibinfo{author}{\bibfnamefont{A.}~\bibnamefont{Roig}}
  (\bibinfo{year}{in preparation}).

\bibitem[{\citenamefont{{Ni\v z\v nansk\'y}}(No change of the $Pna2_1$ crystal
  structure was observed in XRD up to 800\,K)}]{Nizn-priv}
\bibinfo{author}{\bibfnamefont{D.}~\bibnamefont{{Ni\v z\v nansk\'y}}},
  \bibinfo{journal}{private communication.}  (\bibinfo{year}{No change of the
  $Pna2_1$ crystal structure was observed in XRD up to 800\,K}).

\bibitem[{\citenamefont{Shirane et~al.}(2002)\citenamefont{Shirane, Shapiro,
  and Tranquada}}]{Shirane06}
\bibinfo{author}{\bibfnamefont{G.}~\bibnamefont{Shirane}},
  \bibinfo{author}{\bibfnamefont{S.~M.} \bibnamefont{Shapiro}},
  \bibnamefont{and} \bibinfo{author}{\bibfnamefont{J.~M.}
  \bibnamefont{Tranquada}}, \emph{\bibinfo{title}{Neutron Scattering with a
  Triple-Axis Spectrometer}} (\bibinfo{publisher}{Cambridge University Press},
  \bibinfo{year}{2002}), \bibinfo{note}{pp. 36 ff.}

\bibitem[{\citenamefont{Delaire et~al.}(2012)\citenamefont{Delaire, Stone, Ma,
  Huq, Gout, Brown, Wang, and Ren}}]{Delaire12}
\bibinfo{author}{\bibfnamefont{O.}~\bibnamefont{Delaire}},
  \bibinfo{author}{\bibfnamefont{M.~B.} \bibnamefont{Stone}},
  \bibinfo{author}{\bibfnamefont{J.}~\bibnamefont{Ma}},
  \bibinfo{author}{\bibfnamefont{A.}~\bibnamefont{Huq}},
  \bibinfo{author}{\bibfnamefont{D.}~\bibnamefont{Gout}},
  \bibinfo{author}{\bibfnamefont{C.}~\bibnamefont{Brown}},
  \bibinfo{author}{\bibfnamefont{K.~F.} \bibnamefont{Wang}}, \bibnamefont{and}
  \bibinfo{author}{\bibfnamefont{Z.~F.} \bibnamefont{Ren}},
  \bibinfo{journal}{Phys. Rev. B} \textbf{\bibinfo{volume}{85}},
  \bibinfo{pages}{064405} (\bibinfo{year}{2012}).

\bibitem[{\citenamefont{Jeong et~al.}(2012)\citenamefont{Jeong, Goremychkin,
  Guidi, Nakajima, Jeon, Kim, Furukawa, Kim, Lee, Kiryukhin et~al.}}]{Jeong12}
\bibinfo{author}{\bibfnamefont{J.}~\bibnamefont{Jeong}},
  \bibinfo{author}{\bibfnamefont{E.~A.} \bibnamefont{Goremychkin}},
  \bibinfo{author}{\bibfnamefont{T.}~\bibnamefont{Guidi}},
  \bibinfo{author}{\bibfnamefont{K.}~\bibnamefont{Nakajima}},
  \bibinfo{author}{\bibfnamefont{G.~S.} \bibnamefont{Jeon}},
  \bibinfo{author}{\bibfnamefont{S.-A.} \bibnamefont{Kim}},
  \bibinfo{author}{\bibfnamefont{S.}~\bibnamefont{Furukawa}},
  \bibinfo{author}{\bibfnamefont{Y.~B.} \bibnamefont{Kim}},
  \bibinfo{author}{\bibfnamefont{S.}~\bibnamefont{Lee}},
  \bibinfo{author}{\bibfnamefont{V.}~\bibnamefont{Kiryukhin}},
  \bibnamefont{et~al.}, \bibinfo{journal}{Phys. Rev. Lett.}
  \textbf{\bibinfo{volume}{108}}, \bibinfo{pages}{077202}
  (\bibinfo{year}{2012}).

\bibitem[{\citenamefont{Rhyne and Koon}(1978)}]{Rhyne78}
\bibinfo{author}{\bibfnamefont{J.~J.} \bibnamefont{Rhyne}} \bibnamefont{and}
  \bibinfo{author}{\bibfnamefont{N.~C.} \bibnamefont{Koon}},
  \bibinfo{journal}{J. Appl. Phys.} \textbf{\bibinfo{volume}{49}},
  \bibinfo{pages}{2133} (\bibinfo{year}{1978}).

\bibitem[{\citenamefont{Petzelt}(1981)}]{Petzelt81}
\bibinfo{author}{\bibfnamefont{J.}~\bibnamefont{Petzelt}},
  \bibinfo{journal}{Phase Trans.} \textbf{\bibinfo{volume}{2}},
  \bibinfo{pages}{155} (\bibinfo{year}{1981}).

\bibitem[{\citenamefont{Fennie}(2008)}]{Fennie08}
\bibinfo{author}{\bibfnamefont{C.}~\bibnamefont{Fennie}},
  \bibinfo{journal}{Phys. Rev. Lett.} \textbf{\bibinfo{volume}{100}},
  \bibinfo{pages}{167203} (\bibinfo{year}{2008}).

\end{thebibliography}
\end{document}